\documentclass[pra,notitlepage,superscriptaddress,longbibliography,twocolumn]{revtex4-1}
\usepackage[utf8]{inputenc}
\usepackage{graphicx}
\usepackage{amsmath}
\usepackage{amsthm}
\usepackage{bm}
\usepackage{layout}
\usepackage{float}
\usepackage{amsfonts}
\usepackage{amssymb}%
\usepackage[margin=0.75in]{geometry}
\usepackage{color}
\usepackage{soul}
\usepackage{mathtools}

\usepackage{tabularx}
\usepackage{array}

\usepackage[colorlinks=true,citecolor=blue]{hyperref}
\theoremstyle{definition}

\usepackage[capitalise,compress]{cleveref}

\usepackage{physics}
\newcommand{\avg}[1]{\left \langle #1 \right\rangle}
\usepackage{ulem}
\usepackage[subpreambles=false]{standalone}
\usepackage{tikz}
\usepackage[section]{placeins}

\usetikzlibrary{arrows}
\usetikzlibrary{shapes.geometric}

\usepackage{tabularx}
\usepackage[noend]{algpseudocode}

\makeatletter
\newcommand{\multiline}[1]{%
  \begin{tabularx}{\dimexpr\linewidth-\ALG@thistlm}[t]{@{}X@{}}
    #1
  \end{tabularx}
}
\makeatother

\newcolumntype{P}[1]{>{\centering\arraybackslash}p{#1}}

\renewcommand{\epsilon}{\varepsilon}
\renewcommand{\O}[1]{ O\left(#1\right)}

\renewcommand{\emph}[1]{\textit{#1}}

\newcounter{para}

\makeatletter
\newcommand*\bigcdot{\mathpalette\bigcdot@{.5}}
\newcommand*\bigcdot@[2]{\mathbin{\vcenter{\hbox{\scalebox{#2}{$\m@th#1\bullet$}}}}}

\newcommand{\llangle}[1][]{\savebox{\@brx}{\(\m@th{#1\langle}\)}%
  \mathopen{\copy\@brx\kern-0.5\wd\@brx\usebox{\@brx}}}
\newcommand{\rrangle}[1][]{\savebox{\@brx}{\(\m@th{#1\rangle}\)}%
  \mathclose{\copy\@brx\kern-0.5\wd\@brx\usebox{\@brx}}}

\makeatother

\usepackage{array}
\newcolumntype{L}{>{$}l<{$}} 
\newcolumntype{C}{>{$}c<{$}} 
\newcolumntype{R}{>{$}r<{$}} 

\usepackage{xcolor}

\definecolor{borange}{HTML}{BF5700}

\definecolor{darkgreen}{HTML}{009020}

\usepackage{pifont}

\usepackage{mathtools}

\usepackage{xr}
\makeatletter
\newcommand*{\addFileDependency}[1]{
  \typeout{(#1)}
  \@addtofilelist{#1}
  \IfFileExists{#1}{}{\typeout{No file #1.}}
}
\makeatother

\usepackage{tikz}

\usepackage{mdframed}
\newmdenv[topline=false,rightline=false,bottomline=false,linewidth=2pt,linecolor=white!60!black,]{leftborder}

\usepackage{physics}

\usepackage{array}
\newcolumntype{P}[1]{>{\centering\arraybackslash}p{#1}}

\usepackage{bbm}

\usepackage{booktabs}

\makeatletter
\DeclareFontFamily{OMX}{MnSymbolE}{}
\DeclareSymbolFont{MnLargeSymbols}{OMX}{MnSymbolE}{m}{n}
\SetSymbolFont{MnLargeSymbols}{bold}{OMX}{MnSymbolE}{b}{n}
\DeclareFontShape{OMX}{MnSymbolE}{m}{n}{
    <-6>  MnSymbolE5
   <6-7>  MnSymbolE6
   <7-8>  MnSymbolE7
   <8-9>  MnSymbolE8
   <9-10> MnSymbolE9
  <10-12> MnSymbolE10
  <12->   MnSymbolE12
}{}
\DeclareFontShape{OMX}{MnSymbolE}{b}{n}{
    <-6>  MnSymbolE-Bold5
   <6-7>  MnSymbolE-Bold6
   <7-8>  MnSymbolE-Bold7
   <8-9>  MnSymbolE-Bold8
   <9-10> MnSymbolE-Bold9
  <10-12> MnSymbolE-Bold10
  <12->   MnSymbolE-Bold12
}{}

\let\llangle\@undefined
\let\rrangle\@undefined
\DeclareMathDelimiter{\llangle}{\mathopen}%
                     {MnLargeSymbols}{'164}{MnLargeSymbols}{'164}
\DeclareMathDelimiter{\rrangle}{\mathclose}%
                     {MnLargeSymbols}{'171}{MnLargeSymbols}{'171}
\makeatother

\usepackage{float}
\makeatletter
\let\newfloat\newfloat@ltx
\makeatother
\usepackage{algpseudocode}
\usepackage{algorithm}
\makeatletter
\renewcommand{\ALG@name}{Algorithm }
\makeatother
\usepackage{comment}

\begin{document}

\title{Improved Quantum Computation using Operator Backpropagation}
\author{Bryce Fuller}
\affiliation{IBM Quantum, IBM T.J. Watson Research Center, Yorktown Heights, NY 10598, USA}
\author{Minh C. Tran}
\affiliation{IBM Quantum, IBM T.J. Watson Research Center, Yorktown Heights, NY 10598, USA}
\author{Danylo Lykov}
\affiliation{Argonne National Laboratory, Lemont, IL 60439, USA}
\author{Caleb Johnson}
\affiliation{IBM Quantum, IBM T.J. Watson Research Center, Yorktown Heights, NY 10598, USA}
\author{Max Rossmannek}
\affiliation{IBM Quantum, IBM Research Europe -- Zurich, CH-8803 R\"uschlikon, Switzerland}
\author{Ken Xuan Wei}
\affiliation{IBM Quantum, IBM T.J. Watson Research Center, Yorktown Heights, NY 10598, USA}
\author{Andre He}
\affiliation{IBM Quantum, IBM T.J. Watson Research Center, Yorktown Heights, NY 10598, USA}
\author{Youngseok Kim}
\affiliation{IBM Quantum, IBM T.J. Watson Research Center, Yorktown Heights, NY 10598, USA}
\author{DinhDuy Vu}
\affiliation{Department of Physics, Harvard University, Cambridge, MA 02138, USA}
\author{Kunal Sharma}
\affiliation{IBM Quantum, IBM T.J. Watson Research Center, Yorktown Heights, NY 10598, USA}
\author{Yuri Alexeev}
\affiliation{Argonne National Laboratory, Lemont, IL 60439, USA}
\author{Abhinav Kandala}
\affiliation{IBM Quantum, IBM T.J. Watson Research Center, Yorktown Heights, NY 10598, USA}
\author{Antonio Mezzacapo}
\affiliation{IBM Quantum, IBM T.J. Watson Research Center, Yorktown Heights, NY 10598, USA}

\begin{abstract}

    Decoherence of quantum hardware is currently limiting its practical applications. At the same time,
    classical algorithms for simulating quantum circuits have progressed substantially. Here,
    we demonstrate a hybrid framework that integrates classical simulations with quantum hardware to
    improve the computation of an observable's expectation value by reducing the quantum circuit depth. In this framework, a quantum circuit
    is partitioned into two subcircuits: one that describes the backpropagated Heisenberg evolution of an observable, executed on a classical computer, while the other is
    a Schr\"odinger evolution run on quantum processors. The overall effect is to reduce the depths of the circuits executed on quantum devices, trading this with classical overhead and an increased number of circuit executions. We demonstrate the effectiveness of this method on a Hamiltonian simulation
    problem, achieving more accurate expectation value estimates compared to using quantum hardware alone.
    
\end{abstract}
  
\maketitle

\section{Introduction}

Quantum algorithms promise significant advantages over classical methods in many applications such as Hamiltonian simulation~\cite{lloyd_universal_1996} and solving systems of linear equations~\cite{harrow_quantum_2009}.
These advantages can often only be realized for sufficiently large problem instances and typically require coherent implementations of deep quantum circuits.
However, decoherence of current quantum hardware limits their application to short-depth quantum circuits.
Error mitigation  \cite{temme_error_2017,li_efficient_2017,cai2023quantum} has been used to enable accurate calculations on current quantum hardware at a scale beyond brute force classical computation \cite{kim2023evidence}. These methods typically incur a sampling overhead that is exponential in the depth of the circuit.

In response to the cost of error mitigation, recent experimental demonstrations aim to optimize the performance and resource overhead of quantum experiments by leveraging a variety of classical techniques to reduce the depth of executed quantum circuits.
For example, advancements in transpilation have led to more efficient swap routing~\cite{nation2023suppressing,kremer2024practical}, which can result in shallower circuits to execute. 
Tensor networks have been used in conjunction with classical optimizers to improve the accuracy of expectation values for time evolution problems~\cite{m:filippovScalableTensornetworkError2023}. 
Hybrid multi-product formulas~\cite{vazquez2023well, robertson2024tensornetworkenhanceddynamic, robertson2023approximate}  enable performing Hamiltonian time evolution using an ensemble of shallower quantum circuits. 
Recent algorithms for approximate quantum compiling utilize tensor networks and a classical optimizer to compress deep Trotterized time-evolution circuits into shallower approximations~\cite{10.1145/3505181, Rudolph_2024, Ben_Dov_2024, PRXQuantum.2.010342, PhysRevA.101.032310,robertson2023approximate}.
These rapid developments highlight an ongoing need to discover new algorithms to reduce the depth of quantum experiments.

\begin{figure}
\centering
\includegraphics[width=0.48\textwidth]{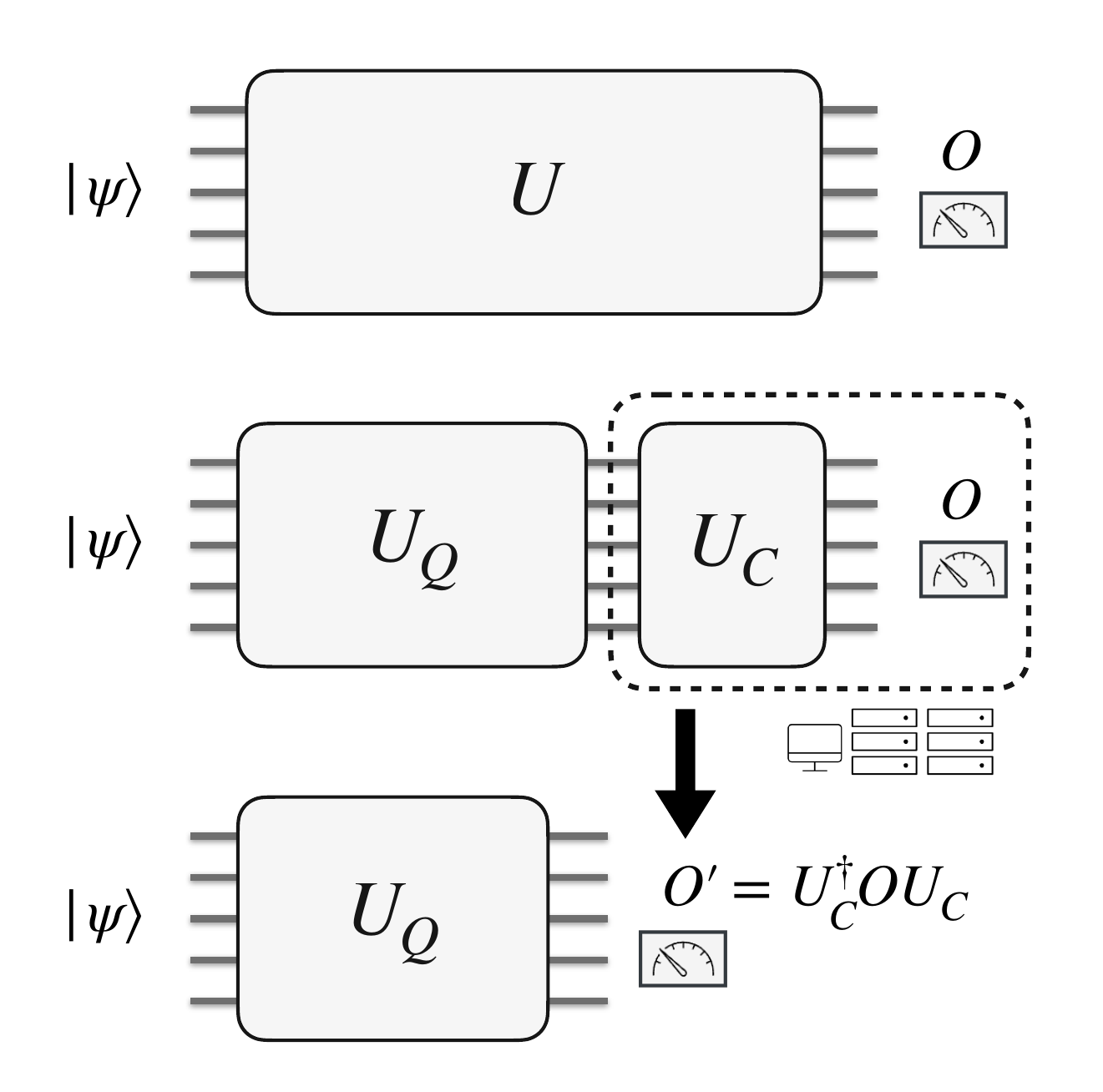}
\caption{\textbf{Operator backpropagation (OBP) framework.} A quantum circuit $U$ is split into two subcircuits $U_C$ and $U_Q$. A classical simulator computes the Pauli decomposition of $O' = U_C^\dag O U_C$, which is then measured on quantum hardware.}
\label{fig:quantum-classical-cut}
\end{figure}

In this manuscript, we introduce a framework to reduce the depth of quantum circuits using classical simulation algorithms based on Clifford perturbation theory (CPT)~\cite{Begusic2023}. We then apply this framework using the Qiskit Addon for operator backpropagation \cite{qiskit-addon-obp} and execute experiments on quantum hardware to observe the reduction in error which can be achieved for a utility scale Hamiltonian time dynamics experiment.
CPT-based algorithms classically compute the expectation value of an observable by \emph{backpropagating} it, i.e. evolving the observable in the Heisenberg picture through the gates of the circuit in reverse order, starting from the last gate of the circuit.
Provided that the circuit consists of a small number of non-Clifford operations, the Pauli decompositions of the backpropagated observable can be computed exactly classically in reasonable runtime. For circuits with a large proportion of non-Clifford gates exact backpropagation grows intractably with circuit depth; however, by allowing for some approximation error a wider range of circuits can be backpropagated in practice.
Such a strategy has been successful in approximating the outcomes of recent utility-scale experiments~\cite{Begusic2024}.
Theoretically, algorithms based on CPT have been shown to be asymptotically efficient for several interesting classes of quantum circuits, including those that are noisy with a random input~\cite{schuster_polynomial_time_2024} or those that consist of random single-qubit operations~\cite{angrisani_classically_2024}.
 
In contrast to these asymptotic analyses, this work uses CPT to approximate the Heisenberg evolution for explicit circuit instances while tracking the accuracy of the calculation using a combination of typical-case error bounds~\cite{schuster_polynomial_time_2024} and the triangle inequality.

In our framework, a quantum circuit is split into two subcircuits~(\cref{fig:quantum-classical-cut}). 
The observable of interest is backpropagated under one of the subcircuits and decomposed as a linear combination of Pauli operators.
The Pauli operators are then measured on the quantum state evolved under the other subcircuit.
In general, the number of Pauli operators grows with the depth of the backpropagated subcircuit. 
Thus, \emph{operator backpropagation} (OBP) allows one to reduce circuit depth in exchange for a classical overhead and an increase in the total number of circuits executed on quantum hardware. Because OBP calculations can become classically expensive and lend themselves to distributed implementation, our method is amenable to be run in quantum-centric supercomputing environments~\cite{alexeev2023quantum}. We describe this framework, including details of the CPT algorithm, in \cref{sec:framework}.
In \cref{sec:experiment}, we demonstrate an application of the framework in improving quantum simulation.
In particular, we show that OBP helps reduce the error in computing the expectation values of observables for circuits with up to 127 qubits and 4896 two qubit gates. 
Hence, given a fixed error tolerance, OBP enables the computation of expectation values for deeper quantum circuits than a purely-quantum approach.
Finally, we discuss open questions and future improvements in \cref{sec:outlook}.

\section{Framework}\label{sec:framework}

Many quantum algorithms rely on measuring operator expectation values with respect to states prepared on quantum devices. 
Specifically, we consider problems of estimating 
\begin{align} 
	\avg{O}_{U\ket{\psi}} 
	\equiv \langle \psi \vert U^\dag O U \vert \psi \rangle \, , \label{eq:expectation-value-problem}
\end{align}
given a quantum state $\ket{\psi}$, a quantum circuit $U$, and an observable $O$.
Without loss of generality, we assume that $O$ is a traceless multi-qubit Pauli operator.
In near-term experiments, the depths of quantum circuits for which the expectation values can be faithfully recovered are constrained by the error of the quantum devices, limiting the size of experiments that can be performed.

Parallel to the advancements in quantum hardware, various classical algorithms have been developed for numerically computing the expectation value in \cref{eq:expectation-value-problem}.
For problems considered in recent experiments, these classical algorithms often perform better or as well as algorithms executed on state-of-the-art quantum devices.

Despite this, the classical complexity of estimating arbitrary expectation values grows exponentially with problem size and, thus, will become out of reach for general classical algorithms even on the largest supercomputers.

To distribute the expectation value problem in \cref{eq:expectation-value-problem} between a quantum device and a classical simulator, we consider a decomposition of $U = U_\text{C}U_{\text{Q}}$ into two subcircuits $U_\text{C}$ and $U_\text{Q}$.
The classical simulator first computes $O' \equiv U_\text{C}^\dag O U_\text{C}$---the version of $O$ evolved through the circuit $U_C^\dagger$.
One then prepares the initial state $\ket{\psi}$ on the quantum hardware, applies the circuit $U_Q$, and measures the expectation value of $O'$.

It is straightforward to verify that the result $\bra{\psi} U_Q^\dag O' U_Q \ket{\psi} = \bra{\psi} U^\dag O U \ket{\psi}$ is the desired expectation value in \cref{eq:expectation-value-problem}.

Standard quantum hardware can measure the expectation values of observables that are diagonal in a local measurement basis. 
To measure the expectation value of $O'$, we require that the classical simulator decomposes it in the Pauli basis, i.e.
\begin{align} 
	O' = \sum_P c_P P \, , \label{eq:pauli-expansion-of-O'}
\end{align}
where $P \in \{I,X, Y, Z\}^{\otimes n}$ are multi-qubit Pauli operators on $n$ qubits and the real numbers $c_P = \Tr(O' P)/2^n$ are the coefficients of the decomposition.
We call the process of approximately evolving $O$ through $U_C^\dagger$ \emph{operator backpropagation} (OBP).
Given the backpropagated observable $O'$, we measure the expectation value of each Pauli in the decomposition and reconstruct the expectation value of $O'$ by
\begin{align} 
	\bra{\psi}U_Q^\dag O' U_Q \ket{\psi} =\sum_P c_P \bra{\psi}U_Q^\dag P U_Q \ket{\psi} \, .
\end{align}
We summarize these steps in \cref{fig:quantum-classical-cut}.

Our framework offers a trade-off between the required circuit depth and the number of circuit executions which are needed to compute $\avg{O}_{U\ket{\psi}}$.
Because the circuit executed on quantum hardware $U_Q$ will be shallower than the original circuit, the resources needed to error mitigate each individual circuit are lowered.
In exchange, the number of distinct circuits which must be executed increases with the number of Pauli operators that compose $O'$ in \cref{eq:pauli-expansion-of-O'}.
In general, the number of Pauli measurements and the error-mitigation overhead both grow exponentially with the depth of $U_C$, thus this framework allows us to trade between these exponentials to optimize the resource requirements and accuracy of quantum hardware experiments. 
The applicability of our framework thus depends on several factors, including the noise profile of the quantum device and details of the problems at hand, e.g. how close the circuits are to Clifford circuits.
In the next section, we discuss  how one can control the truncation error within the classical portion of the OBP algorithm and in \cref{sec:experiment}, we demonstrate an example where this framework allows us to recover expectation values with higher accuracy than experiments which use quantum hardware alone.

\subsection{Truncation error}

Recall that the backpropagated observable $O'$ is written as a linear combination of multi-qubit Pauli operators (cf.~\@\cref{eq:pauli-expansion-of-O'}). Some coefficients $c_P$ in the decomposition  may be small enough that they can be truncated from $O'$ without incurring significant error.
We discuss the estimation of this truncation error in this section.

Suppose that $\mathcal S$ is the set of Paulis upon which $O'$ is supported, $\mathcal T$ is the subset of Paulis which will be truncated, and $\mathcal K$ is the subset of the remaining Paulis which will remain after truncation.
Let $O'_{\mathcal K} = \sum_{P\in\mathcal K} c_P P$ be the truncated version of $O'$.
The difference between them is
\begin{align} 
 	\Delta \equiv O' - O'_{\mathcal K} 
 	= \sum_{P\in\mathcal T} c_P P \, .
 \end{align} 
Using the triangle inequality, the truncation error can be bounded by the $L_1$ norm of the truncated coefficients,
 \begin{align} 
 	 \abs{\bra{\psi_Q} \Delta \ket{\psi_Q}}
 \leq \sum_{P\in\mathcal T} \abs{c_P} \, ,\label{eq:l1-norm-bound}
 \end{align}
 where $\ket{\psi_Q} = U_Q \ket{\psi}$.
This error bound is rigorous, but it can only be saturated if the truncated Pauli operators mutually commute and $\ket{\psi_Q}$ happens to be a common eigenstate of the operators.
Except for these worst cases, the $L_1$ norm of the truncated coefficients will largely overestimate the truncation error.

\begin{figure}
\includegraphics[width=0.45\textwidth]{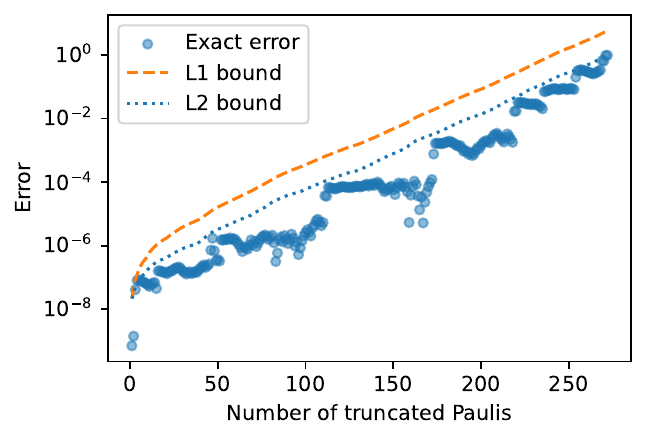}
\caption{
\textbf{Comparison between $L_1$ and $L_2$ error bounds vs. exact error of OBP truncation.}
The exact error $\norm{\Delta}$ from truncating a backpropagated observable (blue circles) and the error bounds based on the $L_1$ norm [\cref{eq:l1-norm-bound}, orange dashed line] and the $L_2$ norm [\cref{eq:l2-norm-bound}, blue dotted line] of the truncated coeffcients.
Here, $U_Q$ and $U_C$ each correspond to 5 Trotter step time evolution circuits for a 12 qubit XY model [\cref{eq:Trotterized_evolution}] on a one-dimensional lattice with closed boundary conditions at Trotter step size $dt = 0.1$.
The initial state is $\ket{00\dots0}$ and the target observable is $Z_1$. 
The total number of Pauli terms in the backpropagated observable without truncation is 271.
}
\label{fig:benchmark_error_bounds}
\end{figure}

Instead, we may use a different estimate that is expected to better capture the truncation error in typical cases.
To motivate the estimate, we first assume that $\ket{\psi_Q}$ were drawn from a 1-design ensemble.
The error in the expectation value follows a distribution with a vanishing mean and a variance given by~\cite{schuster_polynomial_time_2024}
\begin{align} 
	\mathbb E_{\ket{\psi_Q}} \abs{\bra{\psi_Q} \Delta \ket{\psi_Q}}^2 \leq \frac{1}{2^n}\Tr(\Delta^2)=\sum_{P\in\mathcal T} \abs{c_P}^2 \, .
 \end{align} 
Thus, the majority of the drawn $\ket{\psi_Q}$ would result in a truncation error smaller than the $L_2$ norm of the truncated coefficients.
In our problem, $\ket{\psi_Q}$ is deterministic.
But if $U_Q$ is a generic and sufficiently deep circuit, we expect $\bra{\psi_Q} \Delta \ket{\psi_Q}$ to behave as if $\ket{\psi_Q}$ were chosen at random, resulting in an estimate
\begin{align} 
	 \abs{\bra{\psi_Q} \Delta \ket{\psi_Q}} \lesssim \left(\sum_{P\in\mathcal T} \abs{c_P}^2\right)^{1/2} \, .
	 \label{eq:l2-norm-bound}
\end{align}
Compared to the $L_1$-norm bound in \cref{eq:l1-norm-bound}, \cref{eq:l2-norm-bound} may be violated in pathological cases, but is otherwise a better estimate of the truncation error in practice.
We benchmark both approaches in \cref{fig:benchmark_error_bounds} and show that it is indeed the case for even a relatively shallow circuit.

\subsection{Operator Backpropagation via Clifford Perturbation Theory}
\label{sec:framework:CPT}

Our framework requires the decomposition of the backpropagated observable $O'$ in the computational basis.
In this section, we discuss an implementation of such an OBP algorithm based on the Clifford perturbation theory (CPT)~\cite{Begusic2023}.
The key idea is that Clifford circuits can be efficiently simulated by classical computers and we can realize an algorithm using CPT whose complexity scales exponentially with the non-Cliffordness of $U_C$.

The OBP algorithm based on CPT works as follows.
First, we split the subcircuit to be backpropagated, $U_C$, into \emph{slices},
\begin{align}
    U_C &= \prod_{s=1}^S \mathcal{U}_s\equiv \mathcal U_S\dots \mathcal U_2 \mathcal U_1 \, ,
\end{align}
where $S$ denotes the total number of slices and $\mathcal{U}_s$ represents a single slice of $U_C$.
No constraints are made on the depth of the slices and each slice may be supported on all qubits.
The algorithm then proceeds by iterating over the circuit slices \emph{in reverse}, i.e. starting from slice $S$ and ending on slice $1$.
For each slice, all quantum gates are analytically applied to the current operator,
\begin{align}
    O'^{(s)} &= \mathcal{U}_{S-s+1}^\dag O'^{(s-1)}\mathcal{U}_{S-s+1} \, ,
\end{align}
where $s$ denotes the iteration index of the OBP algorithm and $O'^{(0)} = O$ is the original operator whose expectation value we are interested in.

For example, if $O = X_1$ is a Pauli $X$ supported on qubit 1 and $\mathcal{U}_{S}$ is a CNOT gate between qubits $1$ and $2$, $O'^{(1)} = \mathcal{U}_{S}^\dag X_1 \mathcal{U}_{S}$ again contains only one Pauli string $X_1 X_2$.    
However, if $\mathcal{U}_{S}$ were a $T$ gate on qubit 1, $O'^{(1)}$ would instead contain two Pauli operators: $X_1$ and $Y_1$.
In general, the number of Paulis in the backpropagated observable remains the same after applying a Clifford gate and may double after each non-Clifford one.
Therefore, during the course of the OBP algorithm, the number of Pauli terms comprising $O'^{(s)}$ can grow exponentially.

The OBP algorithm based upon CPT is particularly convenient when $U_C$ contains gates that are unitary rotations by small angles.
Such rotations often arise in a Trotterized Hamiltonian simulation.
In such cases, many of the coefficients in the decomposition of $O'^{(s)}$ are small, providing an opportunity to truncate them from the operator.
In our implementation, truncation occurs after the backpropagation of each slice.
This is what motivates the splitting of $U_C$ into slices, as these provide natural stopping points within the OBP algorithm and facilitate the division of a total error budget among the fixed number of truncation steps which occur. 

\begin{figure}[b]
    \centering
    \begin{tikzpicture}

\node at (0.0,  0.0 ) {Paulis};
\node at (0.0, -1.0 ) {\texttt{IIII}};
\node at (0.0, -1.25) {\texttt{IZIZ}};
\node at (0.0, -1.5 ) {\texttt{IXXI}};
\node at (0.0, -1.75) {\texttt{YZIX}};
\node at (0.0, -2.0 ) {$\vdots$};

\node at (2.0,  0.0 ) {ZX calculus};
\node at (1.5, -0.5 ) {\texttt{Z}};
\node at (2.5, -0.5 ) {\texttt{X}};
\node at (1.5, -1.0 ) {\texttt{0000}};
\node at (2.5, -1.0 ) {\texttt{0000}};
\node at (1.5, -1.25) {\texttt{0101}};
\node at (2.5, -1.25) {\texttt{0000}};
\node at (1.5, -1.5 ) {\texttt{0000}};
\node at (2.5, -1.5 ) {\texttt{0110}};
\node at (1.5, -1.75) {\texttt{1100}};
\node at (2.5, -1.75) {\texttt{1001}};
\node at (2.0, -2.0 ) {$\vdots$};

\node at (4.0,  0.0 ) {Address};
\node at (4.0, -1.0 ) {$0$};
\node at (4.0, -1.25) {$80$};
\node at (4.0, -1.5 ) {$6$};
\node at (4.0, -1.75) {$201$};
\node at (4.0, -2.0 ) {$\vdots$};

\draw [->] (4.33, -1.0 ) -- (5.0, -1.0 );
\draw [->] (4.33, -1.25) -- (5.0, -1.25);
\draw [->] (4.33, -1.5 ) -- (5.0, -1.0 );
\draw [->] (4.33, -1.75) -- (5.0, -1.75);

\node at (6.0,  0.0 ) {Nodes};
\node at (5.0, -1.0 ) [anchor=mid west] {node 1: \texttt{0..63}};
\node at (5.0, -1.25) [anchor=mid west] {node 2: \texttt{64..127}};
\node at (5.0, -1.5 ) [anchor=mid west] {node 3: \texttt{128..191}};
\node at (5.0, -1.75) [anchor=mid west] {node 4: \texttt{192..255}};

\end{tikzpicture}
    \caption{
        \textbf{Example of distributing Pauli terms based on their ZX calculus index.}
        From left to right we show how a Pauli term gets encoded in the ZX calculus which then gets interpreted as an address that can be mapped into the address range associated with a given node.
    }
    \label{fig:pauli-dist-example}
\end{figure}

At this point we emphasize once more that the slices are not restricted to be of depth $1$.
For example, when a $X_1X_2$ and $Y_1Y_2$ rotation are applied subsequently on the same pair of qubits 1 and 2, truncating terms between their individual backpropagation may not be desirable, because backpropagating both gates at once may itself result in beneficial cancellation of terms. This choice might be made, a-priori, if one knows that the operator being backpropagated, e.g. $Z_1 + Z_2$, is an approximate symmetry of an $X_1X_2 + Y_1Y_2$ rotation.
Thus, we make no assumption about the depth of each slice.

Truncation of the operator reduces the memory required to store its Pauli decomposition, but it also leads to a challenge in tracking the total truncation error.

Because truncation occurs at different points in the Heisenberg evolution, the orthogonality of truncated components cannot be assumed. Thus one may not use \cref{eq:l1-norm-bound} or \cref{eq:l2-norm-bound} directly.
Instead, one can use the triangle inequality to further upper bound the total truncation error
\begin{align} 
	\epsilon \leq \epsilon_1 + \epsilon_2 + \dots \epsilon_S \, , 
\end{align}
where $\epsilon_s$ is the truncation error, evaluated using either \cref{eq:l1-norm-bound} or \cref{eq:l2-norm-bound}, at the $s$th iteration.

For the implementation of OBP discussed here, a total error budget is specified up front which is divided among each slice of $U_{C}$. At each truncation step, Pauli terms are removed in order of increasing coefficient magnitude until no further terms can be truncated without exceeding the {\it per-slice} error budget. At the end of truncation, any residual error budget is added to the {\it per-slice} error budget of the next slice. The complexity of this procedure scales as $\mathcal{O}(|\mathcal S_s|\log{\mathcal |S_s|})$ where $\mathcal S_s$ is the set of Pauli terms comprising $O'^{(s)}$ prior to truncation.

The implementation of OBP described above will terminate if any of the three following conditions are met:
\begin{itemize}
	\item the observable $O$ has been backpropagated through all circuit slices, $\mathcal{U}_{s}$
	\item after expending the truncation budget, the size of $\mathcal{K}$ is over the user-specified limit
    \item the algorithm exceeded a user-specified runtime limit.
\end{itemize}

The latter two early-termination criteria are motivated by practical constraints. The total number of Paulis in the decomposition of $O'$, i.e. $|\mathcal K|$, is a proxy for both the classical memory requirements as well as the total number of quantum circuits which need to be executed on hardware. It is also possible to terminate based on the minimal number of qubit-wise-commuting Pauli groups, which defines the true number of required circuit executions. However, computing this quantity is NP-Hard~\cite{Yen2020} and approximating it for large $|\mathcal K|$ can be prohibitively expensive.

Multiple approaches can be taken to parallelize the implementation of the OBP algorithm.
The most na\"ive approach would be to simply parallelize the backpropagation of multiple operators of interest but this assumes that one indeed has multiple operators to begin with.
Furthermore, this does not resolve the fundamental limitation imposed by memory being the most likely resource constraint for how many circuit slices can be backpropagated.
Therefore, a multi-node parallelization for the backpropagation of any single operator is desirable.

The OBP implementation based on CPT poses two major difficulties for such a parallelization.
First, when a slice of circuit operations gets backpropagated, each operation has to be applied to each Pauli term.
If the circuit operation commutes with the Pauli term, this is trivial.
Otherwise, if the operation is non-Clifford, the backpropagation results in up to twice the number of Pauli terms, possibly leading to new terms not present previously.
When this is done in parallel on disjoint sets of Pauli terms, the resulting sets may no longer be disjoint.
If one now desires to perform low-magnitude coefficient truncation while adhering to an error budget, the sets of Pauli terms must be de-duplicated such that coefficients of duplicate Pauli terms are properly summed.
Na\"ively, this would involve either an all-to-all or one-to-all communication pattern, posing a severe bottleneck on the parallelization efficiency, and in the latter case, bottlenecking the entire computation by any single node's memory capacity.
Second, agreeing on a truncation threshold over a distributed set of Pauli terms also poses a potential difficulty.
In the following, we are going to propose a solution to overcome both of these limitations.

The first problem can be tackled by realizing that all $4^N$ Pauli terms that span the space of $N$ qubits have a natural ordering.
Using ZX calculus, every Pauli term can be represented via two bitstrings of length $N$.
Each bitstring encodes the presence ($1$) or lack ($0$) of a Z or X Pauli at the given index, respectively~\cite{Coecke2008,Coecke2011}.
This is possible due to the relations of the identity and three Pauli matrices.
Therefore, in this representation, the four possible combinations of two bits encode the following: $(0,0) \rightarrow I$, $(0,1) \rightarrow Z$, $(1,0) \rightarrow X$, $(1,1) \rightarrow Y$.

Further, we can concatenate the bistrings encoding the Z and X Paulis to form a single bitstring of length $2N$.
This bitstring serves as a unique identifier for every possible Pauli term and provides a natural order to the entire set.
Thus, given any Pauli term, its bitstring can be used as an address to quickly determine its position in the larger set.

When distributing Pauli terms across $R$ compute nodes, the entire address space can be partitioned it into $R$ intervals, associated to the different nodes. This distribution scheme ensures that  for any Pauli address, the corresponding node whose interval contains this Pauli can be determined in constant time.
This procedure is visualized in Fig.~\@\ref{fig:pauli-dist-example} for a specific case involving 4 qubits and a selection of Pauli terms which are to be assigned to one of the four compute nodes.
From left to right, the figure shows how a Pauli gets encoded in the ZX calculus which then gets interpreted as an address that can be mapped into the address range associated with a given node.

The scheme presented above addresses the challenge of distributed deduplication and requires each node to exchange at most $R$ messages. In \cref{sec:load-balancing}, we further show how the partitioning of Pauli addresses can be efficiently updated in order to balance the number of Paulis within each node's interval, requiring $\mathcal{O}(R)$ total messages to be passed.

Now that a distributed storage system for the Pauli terms has been devised, backpropagation of any circuit slice can be performed in parallel on disjoint sets of Pauli terms.
If we wish to truncate Pauli terms with low-magnitude coefficients, this can also be done in parallel but care must be taken not to exceed any specified error budget. 
To this end, every node can in parallel determine the smallest and largest coefficient magnitude and communicate these values to a master node.
Upon receiving all lower and upper bounds, the master node can propose a truncation threshold and broadcast this to all nodes.
Each node can then compute its own truncation error for the proposed threshold and send this information back to the main rank. 
By iterating on this procedure, a binary search for an agreeable truncation threshold can be performed on a distributed set of Pauli terms and coefficients.
Therefore, $\mathcal{O}(R\ \log(|\mathcal{S}|))$ broadcast messages are required to determine the truncation threshold, where $\mathcal{S}$ is the set of Paulis upon which the backpropagated observable is supported, prior to truncation.

\section{Experiment}\label{sec:experiment}

\begin{figure*}
    \centering
    \includegraphics[width=0.95\linewidth]{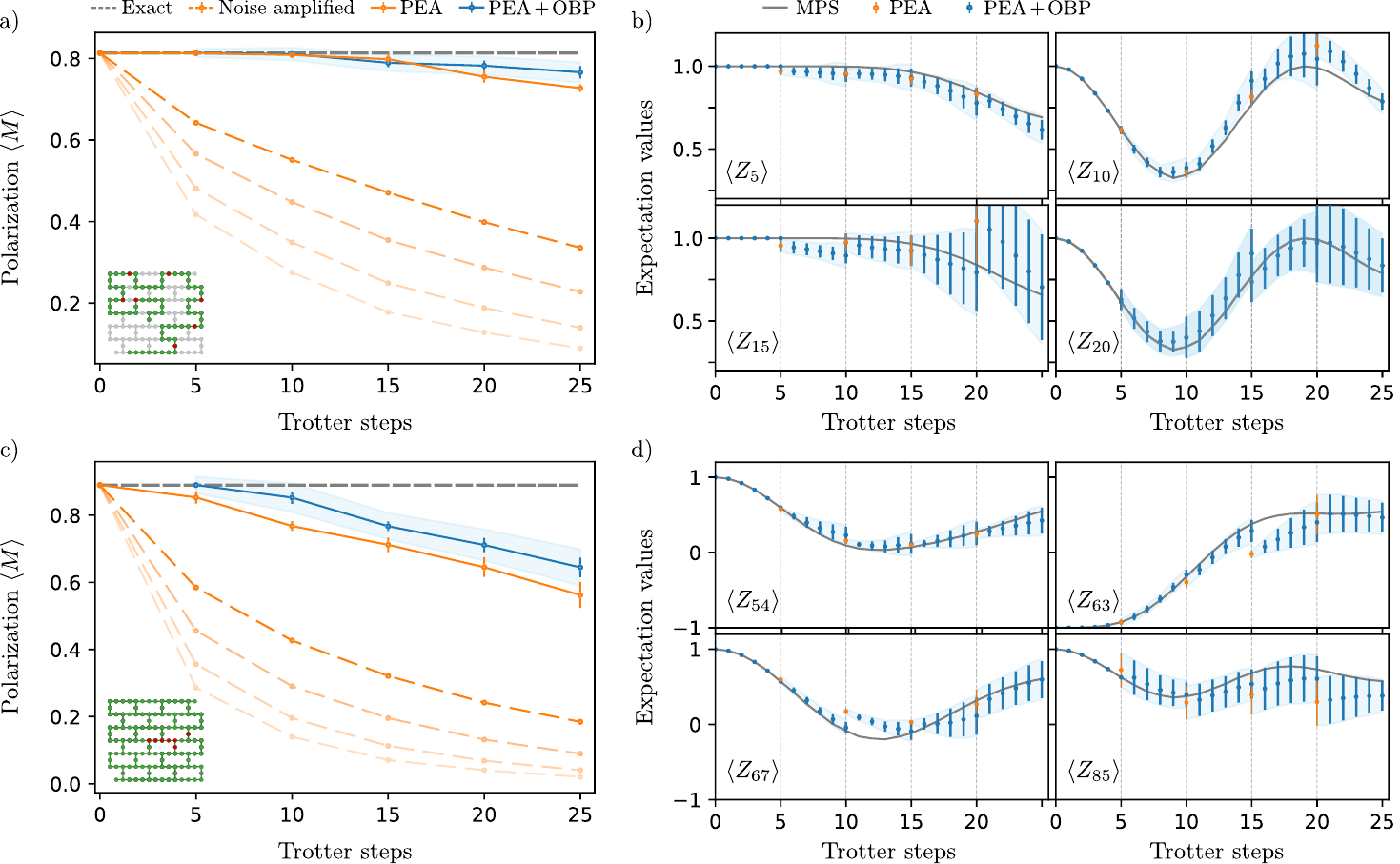}
    \caption{ \textbf{OBP experiments with 75 and 127-qubit spin models.}
        Benchmarking the OBP framework in the simulation of the one-dimensional XY model of 75 spins [panels a) and b)] and the two-dimensional XY model of 127 spins [panels c) and d)].
        a/c) Expectation of the polarization $M$ at different time steps. The polarization is a conserved quantity (dashed line) under the dynamics of the XY model.
        Due to noise, the experimental signal decays with the depth of the circuit and can be partially recovered using error mitigation. The signals at different noise amplification (by a factor of 1, 1.5, 2.25 and 3, indicated by bolder to more transparent dashed orange lines, respectively) are extrapolated to obtain the PEA estimate (solid orange lines). 
        Using OBP with 5 Trotter steps backpropagated, the polarization can be measured to a higher accuracy in deep circuits (blue lines).  
        The insets highlight the qubits of {\texttt{ibm\_kyiv}} used to represent the spins. 
        The qubits are initialized in either $\ket{0}$ (green circles) or $\ket{1}$ (red circles).
        b/d) Dynamics of several individual $Z_i$ under the XY model. 
        The vertical dashed lines indicate Trotter steps at which the expectation values are measured. The orange scatter points indicate results from measurement at 5, 15, and 20 Trotter steps and applying PEA without OBP. The OBP framework helps recover the dynamics of intermediate time values (blue scatter points) from these coarse measurement data. 
        The results agree with the reference values (solid gray lines) obtained via an MPS simulation. 
        All error bars shown were obtained through bootstrapping with 100 batches, and are shown at a 2-$\sigma$ confidence. The shaded blue region represents the additional $L_2$ error bound due to the classical approximation of the backpropagated observable. 
    }
    
    \label{fig:hw_OBP_vs_no_OBP}
\end{figure*}

In this section, we demonstrate the operator backpropagation technique on a Hamiltonian simulation experiment.
In particular, we consider the simulation of the XY model with nearest-neighbor couplings
\begin{equation}
    H = \sum_{i,j\in \mathcal E(\Lambda)} J\left( X_i X_j + Y_i Y_j\right) + h \sum_{i\in\Lambda} Z_i \, ,\label{eq:XY-model}
\end{equation}
where $X_i, Y_i, Z_i$ are the Pauli operators supported on site $i$ and $\mathcal E(\Lambda)$ is the set of edges of a $D$-dimensional regular lattice $\Lambda$, e.g. a one-dimensional chain or a two-dimensional heavy-hex lattice.
For all experiments performed, we consider this model where $J = 1$, $h=0$, and we are interested in estimating the polarization $M \equiv \frac{1}{n}\sum_i Z_i$.

To simulate $e^{-it H}$, we Trotterize the time evolution and approximate it with $U(\tau)^{t/\tau}$, where $\tau$ is the Trotter step size, $t$ is taken to be an integer multiple of $\tau$, and 

\begin{equation}
    U(\tau) = 
    e^{-ih\tau \sum_i Z_i}
    \prod_{i,j\in \mathcal E(\Lambda)}
    e^{-iJ\tau (Y_i Y_j + X_i X_j)}\, .\label{eq:Trotterized_evolution}
\end{equation}
Since this Trotterization preserves the $U(1)$ symmetry of the XY model, the expectation value of $M$ is exactly computable and the error in the measured expectation can be used as a proxy for the error of the simulation. Although $M$ is a conserved quantity of our Trotter circuits, recovering the dynamics of individual $Z_i$ operators is not classically efficient in general when $\Lambda$ is two-dimensional. Using OBP we recover these individual expectation values as well and compare against values obtained by a matrix product state (MPS) calculation. 

All experiments were error mitigated using a combinations of zero-noise extrapolation (ZNE)~\cite{temme_error_2017,li_efficient_2017,kim_scalable_2023,kandala_error_2019} via probabilistic error amplification (PEA)~\cite{li_efficient_2017,mcdonough_automated_2022,mari_extending_2021,ferracin_efficiently_2024, Endo_2018} as well as twirled readout error exctinction (TREX)~\cite{van_den_berg_model-free_2022}. The noise learning and noise amplification procedures used to apply PEA were
performed as in \cite{kim2023evidence}. In \cref{sec:error_mitigation} we discuss further the details of the error mitigation used.

To benchmark the framework, we first apply it to the simulation of the exactly solvable one-dimensional XY model of 75 spins.
We observe that as circuit depth increases, experiments leveraging OBP obtain lower error for estimates of polarization when the total number of circuit executions are held constant. In addition, we run a larger experiment of a two-dimensional XY model on heavy-hex lattice of 127 spins. 
This system is not exactly solvable and other two-dimensional geometries are known to exhibit interesting phenomena such as topological phase transitions \cite{dingpt2DXY}. 
We observe that for all Trotter depths considered, experiments which estimate the polarization using OBP obtain a reduction in error relative to experiments which do not use OBP. 

For both the one and two-dimensional spin models, the spins are mapped to a subset of nearest-neighboring qubits on the coupling graph of {\texttt{ibm\_kyiv}}, one of IBM Quantum's superconducting quantum processors. We fix the length of each Trotter step to be $\tau = 0.05$ and consider the expectation value of the polarization, i.e.

\begin{equation}
\langle M \rangle_k = \bra{\psi}U^\dagger(\tau)^{k} M U(\tau)^{k}\ket{\psi},
\label{eq:noisy_expval}
\end{equation}
at different numbers of Trotter steps $k$, with and without using OBP to reduce the circuit depth by 5 Trotter steps. 
We chose this value by fixing an $L_2$ error budget for the classical approximation of $M$, fixing a maximum of 10 qubit-wise commuting Pauli groups, and then selecting the largest whole number of Trotter steps that could be backpropagated within these constraints. In \cref{sec:numerics_details} we further discuss the details of the classical OBP calculations.

For each OBP computation, the polarization $M \equiv \frac{1}{n}\sum_i Z_i$ is processed for five Trotter steps by backpropagating each $Z$ operator independently using an $L_2$ error budget of $0.01$ and $0.025$ for the one and two dimensional models, respectively. In each case, the error budgets were distributed unevenly across the circuit slices, withholding a majority of the error budget for the final slice. For both models, the set of Paulis supporting all backpropagated observables are collected into 8 qubit-wise commuting groups, each of which is independently evaluated on a QPU. 

Additionally, the Pauli operators needed to reconstruct one through four Trotter steps are a subset of those needed to reconstruct five steps. Thus, we can re-use our measurement data to recover the dynamics of both XY models for all $k\in[0,25]$, despite the fact that we only execute circuits on hardware where $k$ is an integer multiple of 5.

The results shown for all experiments are obtained by bootstrapping the measurements into 100 batches and then taking the mean across all postprocessed batches. The error bars are plotted with a $2\sigma$ confidence and data which leverages OBP is plotted with a shaded region which includes the statistical error of the experiment as well as the $L_2$ error budget which was allocated during backpropagation. 
{See \cref{sec:error_mitigation} for more information on the error mitigation used for this experiment.}

In \cref{fig:hw_OBP_vs_no_OBP}a, we plot the expectation value of the polarization, i.e. \cref{eq:noisy_expval} for a one dimensional XY model of 75 spins. The deepest circuit executed for this experiment is 25 Trotter steps, which requires 1924 two-qubit gates and a two-qubit circuit depth of 52. The initial state $\ket{\psi_0}$ is initialized to $\ket{00\dots0}$, except for seven evenly spaced qubits which are initialized to $\ket{1}$.
Under the dynamics of the XY model, these ``excitations" (qubits initialized to $\ket{1}$) will spread to other sites on the lattice. We reconstruct the expectation value of all local $Z$ operators with and without OBP and compare their mean to the known reference value of $\frac{75-2*7}{75} \approx 0.813$. The experiments performed with and without OBP each use a total $262144$ circuit executions for every $k$. We observe that after 15 Trotter steps, the experiments leveraging OBP achieve a statistically significant reduction in error when estimating $\langle M \rangle_k$. 
Although the OBP framework will, in general, introduce an overhead in the number of quantum circuit executions (shots), these results highlight that even for a fixed budget of circuit executions, one can reduce the error of a quantum simulation by reducing circuit depth via OBP.

{
In \cref{fig:hw_OBP_vs_no_OBP}b, we highlight another important feature of the OBP framework: the ability to recover the dynamics at intermediate times from coarse measurement data.
Specifically, from the measurement after only 5, 10, 15, 20, and 25 Trotter steps, we can reconstruct using OBP the expectation values of  observables at all Trotter steps from 0 to 25.
We plot the dynamics of several such observables $Z_i$ in \cref{fig:hw_OBP_vs_no_OBP}b and compare against reference values obtained from MPS simulations. The $Z_i$ displayed are chosen with $i$ located near the initial excitations (red circles in the inset of \cref{fig:hw_OBP_vs_no_OBP}a) in order to observe strong dynamical response to the excitations. We note that our estimates of the individual $Z_i$ dynamics are more impacted by noise than our estimates of $\langle M \rangle_k$, which is explained by a concentration about the mean arising from the averaging over all $Z_i$. Nevertheless, we observe qualitative agreement between the fine-grained dynamics reconstructed via OBP and those obtained via a MPS simulation. 
}

In addition to the exactly solvable XY spin chain, we also consider a two-dimensional XY model of 127 spins defined on a heavy-hex graph given by the qubit connectivity of the {\texttt{ibm\_kyiv}} QPU. The largest circuit executed in this experiment is $25$ Trotter steps, which requires 4896 two-qubit gates and a two-qubit depth of 102. Because this model is two-dimensional, the resulting Trotter circuits are approximately twice as deep and contain $254\%$ as many two-qubit gates as the Trotter circuits for the one-dimensional model. These circuits are more strongly impacted by noise, and the light-cones of operator expectation values spread more rapidly across the system. In \cref{sec:circuit-synthesis} we detail how the Trotter circuits in this experiment were synthesized.

In \cref{fig:hw_OBP_vs_no_OBP}c and \cref{fig:hw_OBP_vs_no_OBP}d, we plot $\langle M \rangle_k$ for the two-dimensional XY model as well as the expectation values for a handful of local $Z$ observables, which we compare with reference values obtained through MPS simulations. The MPS calculations are discussed in more detail in \cref{sec:mps_simulations}. Similar to the one-dimensional experiment, we initialize $\ket{\psi_0}$ to $\ket{00...0}$ apart from seven excitations which are placed near the center of the lattice. We reconstruct the expectation values of all local Z operators with and without OBP and compare the mean with the known reference value of $\frac{127-2*7}{127} \approx 0.89$. For these experiments, each group of qubit-wise commuting Paulis estimated on hardware is allocated $32768$ shots, resulting in a total of $262144$ ($32768$) shots for estimates of $\langle M \rangle_k$ with (without) OBP. 
We observe that for all number of Trotter steps, estimates of $\langle M \rangle_k$ which use OBP to reduce the circuit depth by 5 steps obtain a statistically significant decrease in error relative to experiments which do not make use of OBP.

\section{Conclusion and Outlook}
\label{sec:outlook}
In this paper, we have introduced the OBP framework for improving the quantum computation of expectation values for local observables on pre-fault tolerant hardware
and we have demonstrated on quantum hardware how OBP can reduce the error of a Hamiltonian time dynamics experiment. This framework reduces the depth of quantum circuits run on quantum hardware, limiting the impact of errors and the cost of error mitigation in exchange for an increase in the number of experiments and an additional classical overhead.
In the context of Hamiltonian simulation, OBP reduces the error in approximating the dynamics of quantum systems, allowing longer-time simulation compared to purely quantum approaches.

An important feature of the OBP framework is the ability to reconstruct dynamics at intermediate times from coarse measurement data. 
For example, in our experiment, we only measure the state after 5, 10, 15, 20, and 25 Trotter steps, but the expectation value of the observable can be constructed for all Trotter steps between 0 and 25. 
Additionally, the same expectation value can be computed in multiple ways: for example, the expectation value after 15 Trotter steps can also be computed from the measurement after 10 Trotter steps using 5 backpropagated Trotter steps. 
In principle, these multiple ways to estimate the same quantity, each suffering from a different set of errors, may be used to obtain a more accurate estimate of the expectation value. 
Exploring such an error-mitigation scheme is an interesting future direction.

The OBP framework is convenient whenever measuring the backpropagated observable requires a manageable number of quantum circuit executions. 
Generally, this requirement puts a constraint on the depth of the circuit $U_C$ through which an observable can be backpropagated. If one fixes an error budget and a maximum overhead in the number of observable measurements, then the reduction in circuit depth which can be achieved will depend strongly on the circuit which one attempts to backpropagate. The OBP framework will show the most promise for problems where the size of one's backpropagated observable grows slowly with the depth of $U_C$. This can be achieved in situations where, for example, $U_C$ consists of few non-Clifford operations or the dynamics of a quantum systems experience periodicity at stroboscopic times, at which point the backpropated observable may be decomposed into a small number of Pauli operators. 
Identifying such use cases where backpropagation yields substantial depth reduction while the full experiment remains classically hard to simulate is an interesting direction for future work.

\emph{Note added:} During the preparation of this manuscript, we learned of a complementary work by Faehrmann \emph{et al.}~\cite{faehrmann_short-time_2024}.
The authors proposed to backpropagate an observable through a short-time evolution of a Hamiltonian using the truncated Taylor series. 
In general, one can instead construct a Trotterized approximation to the dynamics and apply the OBP framework detailed in this manuscript.  
Comparing the efficiency of the two approaches and identifying applications where one can be more beneficial than the other is of great practical interest, but is outside the scope of the current manuscript.

\begin{acknowledgments}
 
This material is based upon work supported by the U.S. Department of Energy, Office of Science, and National Quantum Information Science Research Centers. Y.A. and D.L. acknowledge 
support from the U.S. Department of Energy, Office of Science, under contract DE-AC02-06CH11357 at Argonne National Laboratory. We would also like to acknowledge informative discussions with Petar Jurcevic, Oles Shtanko, Bibek Pokharel, Andrew Eddins, Zlatko Minev, Conrad Haupt, and Daniel Egger. 

\end{acknowledgments}

\appendix

\section{Load balancing of parallelized operator backpropagation}\label{sec:load-balancing}
In order to parallelize the backpropagation of a single observable across multiple nodes, a scheme is needed to deduplicate Pauli operators prior to truncation. One solution is to leverage the ZX calculus, which associates each $N$-qubit Pauli in a backpropagated observable to a unique bit string of length $2N$ which we denote as the Pauli address. For $R$ nodes, we can then split the interval of $[0, 4^N -1 ]$ into $R$ partitions and assign each interval to a different node. 
This allows for a distributed deduplication of Paulis, because after backpropagating one slice, each node knows which node each new Pauli operator should be sent to, requiring at most one message to be sent and received between each of the other $R-1$ nodes. However, this scheme does not fully circumvent the risk of becoming memory bottle-necked by a single node unless it can also update each node's interval to balance the number of Paulis within each interval. Here we present an efficient procedure for updating the intervals of all nodes.

Let the partition of Pauli addresses be given by an increasing sequence of $R+1$ integers \{$\mathcal{B}_i$\}, where $\mathcal{B}_0 = 0$, $ \mathcal{B}_{R} =  4^N$ and  $\mathcal{B}_i < \mathcal{B}_{i+1}$. Let $[ \mathcal{B}_{r}, \mathcal{B}_{r+1} )$ be the interval of Pauli addresses assigned to node $r$, and $\mathcal{S}^{(r)}$ be the subset of Paulis for some backpropagated observable which fall within this interval. 

Algorithm \ref{alg:load-balance} shows a distributed subroutine for updating \{$\mathcal{B}_i$\} after deduplication has been completed. The algorithm begins with each node broadcasting the number of Paulis, $|\mathcal{S}^{(r)}|$, which requires $2R$ total messages to be passed in all-to-one and then one-to-all fashion. Once all nodes know $|\mathcal{S}| = \sum_r|\mathcal{S}^{(r)}|$, the algorithm proceeds by iterating over nodes in increasing order. At iteration $r$, one can assume the previous nodes have already had their interval boundaries adjusted such that they contain $|\mathcal{S}|/R$ addresses each. In order to balance the load of node $r$, $\mathcal{B}'_{r+1}$ must be updated. Depending on whether this upper boundary needs to be raised or lowered, this step will require a bisection search to be done by node either $r$ or node $r+1$, requiring one or two messages to be passed, respectively. A total of $R-1$ boundaries must be updated in this fashion, requiring at most $2(R - 1)$ messages to be passed. The algorithm finishes with all nodes broadcasting their updated boundaries \{$\mathcal{B}_i$\}, once again requiring $2R$ messages to be passed. Thus the total communication complexity of updating the partitions is $\mathcal{O}(R)$ and the time complexity is $\mathcal{O}(R\log{|\mathcal{S}|})$. 

\begin{algorithm}\label{alg:load-balance}
\caption{Distributed Pauli Partition Resizing}
\begin{algorithmic}[1]
\State Let $\mathcal{S}$ be a subset of $N$-qubit Paulis, $|\mathcal{S}| \ll 4^N$
\State Let $\mathcal{S}^{(r)}$ be the subset of $\mathcal{S}$ with addresses in the interval  $[ \mathcal{B}_{r}, \mathcal{B}_{r+1} )$
\State $L \gets \sum_r |\mathcal{S}^{(r)}|$ \Comment{$2R$ messages}
\State $r \gets 0$
\While{$r < R$}
\If{$|\mathcal{S}^{(r)}| > \frac{L}{R}$}
    \State \multiline{Node $r$ computes $\mathcal{B}'_{r+1}$ s.t. $ [ \mathcal{B}_{r}, \mathcal{B}'_{r+1} )$ \\ contains $L/R$ addresses.}
    \State $\mathcal{B}_{r+1} \gets \mathcal{B}'_{r+1}$ \Comment{1 message}
\ElsIf{$|\mathcal{S}^{(r)}| < \frac{L}{R}$}
    \State $\Delta_r \gets  \frac{L}{R} - |\mathcal{S}^{(r)}| $ \Comment{1 message}
     \State \multiline{Node $r+1$ computes new threshold $\mathcal{B}'_{r+1}$ \\ s.t. $ [ \mathcal{B}_{r+1}, \mathcal{B}'_{r+1} )$ contains $\Delta_r$ addresses.}
    \State $\mathcal{B}_{r+1} \gets \mathcal{B}'_{r+1}$  \Comment{1 message}
\EndIf
\EndWhile
\State $\mathcal{B}_{0}, ... \mathcal{B}_{r}$ are broadcast. \Comment{$2R$ messages}
\end{algorithmic}
\end{algorithm}

\section{XY model Trotter circuit synthesis}\label{sec:circuit-synthesis}

The Trotterized time evolution circuits considered in this manuscript are of the form 

\begin{equation}
    U(\tau) = 
    e^{-ih\tau \sum_i Z_i}
    \prod_{i,j\in \mathcal E(\Lambda)}
    e^{-iJ\tau (Y_i Y_j + X_i X_j)}\, ,\label{eq:time_ev_a0}
\end{equation}
where $J = 1$ and $h = 0$ and $\mathcal E(\Lambda)$ is taken to be the edge set of the lattice $\Lambda$, which can be embedded into a heavy-hex lattice. An initial synthesis of these circuits might begin by partitioning $\mathcal E(\Lambda)$ into disjoint edge sets, and using these sets to apply the rotations $e^{-iJ\tau (Y Y + X X)}$ in parallel. A heavy-hex lattice can be partitioned into no less than three disjoint sets, which we will label as $R$, $G$, $B$. Thus we can rewrite \cref{eq:time_ev_a0} as

\begin{equation}
    U(\tau) = U_R U_G U_B,
\end{equation}
where the dependence on $\tau$ is dropped to reduce visual clutter in subsequent equations
and
\begin{equation}
\begin{aligned}
    U_C = \prod_{i,j\in C} e^{-iJ\tau (Y_i Y_j+X_i X_j)},\quad C \in [R,G,B].
\end{aligned}
\end{equation}

Since an arbitrary rotation under $XX + YY$ can be synthesized using two non-parameterized two-qubit gates,  $U(\tau)$ can be synthesized with a two-qubit depth of 6. For one dimensional $\mathcal E(\Lambda)$ which can be partitioned into only two disjoint edge sets the analogous two qubit depth is 4. 

These circuits can be simplified further by observing that adjacent operations in neighboring Trotter steps can be combined for further depth reduction. If we choose to reverse the order of interactions in every other Trotter step, we can ensure a situation where such a simplification is possible:
\begin{equation}
\begin{aligned}
    U(\tau)^{(2k+1)} &= \left(U_R U_G U_B U_B U_G U_R\right)^{\lfloor{k/2}\rfloor}\left(U_R U_G U_B\right)^{k\mathbin{\%}2}, 
\end{aligned}
\end{equation}
where $\mathbin{\%}$ indicates the modulus operator. This alternating interaction order 
effectively realizes the second-order Trotterization and 
allows us to simplify two consecutive $XX + YY$ rotations between every pair of adjacent Trotter steps, leading to a reduction in the total two-qubit depth of $2(k-1)$ for a circuit with $k$ Trotter steps. Thus, the final circuit depth for a circuit of $k$ Trotter steps is $4k + 2$ or $2k+2$ when $\mathcal E(\Lambda)$ can be partitioned into three or two disjoint edge sets, respectively.

\section{Operator backpropagation via Tensor Network Contractions}

In our framework, we discussed OBP using Clifford Perturbation Theory. 
In this section, we discuss another approach to OBP using tensor networks.
In particular, we
show how to form a tensor network that represents the coefficients of an
operator evolved by a quantum circuit and sample for large
coefficients from the tensor network representation.

Tensor network (TN) contraction cost does not depend on the values of the tensors,
in case the contraction is exact. This fact allows us to simulate the evolution for any
Trotter step size $\tau$, which is a useful advantage over the CPT simulation. In TN-based simulations, the main limit limiting for increasing $\tau$ is the number of significant Paulis
in the resulting state. In contrast, for CPT, it is the maximum number of significant Paulis at each evolution depth.

\paragraph{Tensor representation of operators in the Pauli basis}
Recall that in  \cref{eq:pauli-expansion-of-O'}, we decompose a general operator in the Pauli basis  as
$O = \sum_P c_P P$
where the sum is over $4^n$ Pauli strings $P\in \{I,X,Y,Z\}^{\otimes n}$ and $c_P$ are the coefficients. 
Explicitly labeling each Pauli string by $n$ indices, i.e. $P = \sigma^{i_1}\otimes \sigma^{i_2}\otimes \dots \otimes \sigma^{i_n}$
for $i_1,\dots,i_n \in \{I,X,Y,Z\}$,
the coefficient tensor $c_{i_1,\dots,i_n}$ is effectively a rank-$n$ tensor with $n$ indices, each of dimension 4.
Generally, at large $n$, it takes a prohibitively large memory space to store the coefficient tensor.
However, in certain cases, the coefficient tensors may admit a product form and can be stored efficiently.
For example, the coefficient tensor for a single Pauli string is simply the outer product of $n$ rank-1 tensors.

\paragraph{Evolving the operator in the Heisenberg picture}
Given observable $O$, we can backpropagate it through some unitary $U$ by evolving it in the Heisenberg picture: 
\begin{equation}
    O' = U^\dagger O U,
    \label{eq:heisenberg_pic}.
\end{equation}
Given a tensor representation of the operator $O$ and of the channel $U^\dag \cdot U$ in the
Pauli basis, we can contract the tensors according to \cref{eq:heisenberg_pic} to find the TN representation of the evolved operator $O$.

To find the tensor representation of the channel $U^\dag \cdot U$, we note that the action of a gate $U$ on an observable is a
linear operation. Thus, we
can represent the action of $U$ as a matrix acting on the tensor representation of $O$.
For example, for a 1-qubit observable the action of the gates $X$ and $H$ are represented as the following matrices:
\begin{equation*}
    \mathcal U_X=
    \begin{pmatrix*}[r]
        1 & \ \ 0 &  0 & 0 \\
        0 & 1 &  0 & 0 \\
        0 & 0 & -1 & 0 \\
        0 & 0 & 0  & -1 \\
    \end{pmatrix*},
    \quad
    \mathcal U_H =
    \begin{pmatrix*}[r]
        1 & \ \ 0 &  0 &  \ \ 0 \\
        0 & 0 &  0 & 1 \\
        0 & 0 & -1 & 0 \\
        0 & 1 & 0  & 0 \\
    \end{pmatrix*}
    .
\end{equation*}
Generally, since we are expanding operators in the Pauli basis, the tensor representation of a channel is simply the Pauli transfer matrix of the channel, i.e. a matrix $\mathcal U$ such that
\begin{align}
U^\dag P U = \sum_{P'} {\mathcal U}_{P',P} P',
\end{align}
for all Pauli strings $P$.
Again, labeling each Pauli string by $n$ indices, $\mathcal U$ is essentially a rank-$2n$ tensor.
The tensor representation of the backpropagated operator $O'$ can, in principle, be computed through a tensor contraction:
\begin{align}
    c'_{P'} = \sum_{P} \mathcal U_{P',P}c_P. 
\end{align}
In practice, we do not contract the tensors according to the order of the gates in the circuit. Instead, we construct the full tensor network that represents the final backpropagated observable and optimize the contraction order as described below. 
Note that the optimal contraction order may be different from the order that the gates are applied. For example, given a shallow circuit of a large number of qubits, contracting along the qubit dimension would result in smaller memory footprint and faster computation.

Given a circuit $U$ consisting of one- and two-qubit gates, one may construct the tensor representation of individual gates and contract appropriate indices to form the tensor network that represents the circuit $U$.
Note that the tensor that represents a $k$-qubit gate is simply an outer product between a rank-$2k$ tensor and $n-k$ rank-$2$ identity tensors. Therefore, the tensor network for $U$ can be constructed efficiently. 

\paragraph{Lightcone simplification}

When backpropagating observables, we can optimize our calculation by observing that gates which commute with our observable have no effect on the OBP computation. For sparse observables this allows us to discard gates from our circuit which act on identity terms. As we evolve further backwards through layers of operations, the support of our operator will generally spread out limiting the benefit of this technique for deep circuits. The set of circuit operations which are causally connected to the final value of an observable is known as the observable's lightcone, and can be inferred from the circuit's structure to prevent unnecessary calculation. This procedure mirrors similar techniques used for light-cone optimization in TN-based quantum circuit state simulation~\cite{Galda_Transferability, Shaydulin_Symmetry, LightconeTN}.

\paragraph{Sampling}

The goal of our simulation is to obtain a subset of Pauli terms from our backpropagated observable which have the largest coefficient magnitudes. Once a TN representation of our backpropagated observable is given, all Pauli coefficients can be simultaneously calculated by contracting all of the TN indices except for open indices. However, this procedure produces a dense tensor of size $4^n$ which is prohibitively large. 
Alternatively, given a bitstring which uniquely indexes one of the $4^n$ Paulis in our decomposition, we can query the coefficient of that Pauli without incurring the same prohibitive scaling, however, it is not possible to predict a-priori what Paulis will have high magnitude coefficients. Instead, our goal is to sample bitstrings whose corresponding Pauli terms have high magnitude coefficients and approximate the desired set of Pauli terms probabilistically. This approach is commonly referred to as importance sampling. We note that similar problems occur when sampling from tensor networks which represent quantum states. In that case, a TN encodes the probability of observing different measurement outcomes, and one is often concerned with computing high probability bitstrings. For a review on different approaches to sample from tensor networks see~\cite{Bravyi_Marginals}.
Note that, to reduce the cost, in our approach we use an unmodified TN which represents the real coefficients rather than their absolute values. 
While the marginals calculated in this approach do not strictly correspond to the marginals of the distribution we would like to sample from, in practice, sampling from this TN still results in the same important bitstrings as in the CPT simulations.

\begin{algorithm}[t]
\caption{Distributed sampling from a TN\label{alg:largest_sampling}}
\begin{algorithmic}[1]
\State Let $\mathcal{T}$ be the input tensor network
\State Let $J$ be the set of TN indices that correspond to sample bits.
\State Let $M$ be a batch size parameter, $0 < M \leq |J|$.
\State Let $r$ be the MPI rank (worker ID).
\State
\State Set $B$ be the list of bitstrings, to be populated during the algorithm\label{alg:line:bitstrings}

\State $F \gets \emptyset$ free indices. Start with full TN contract for normalization.
\While{$|B[0]| < |J|$} \Comment{{\scriptsize{ Until all output indices are sampled}}}
    \State $P_r \gets \texttt{find\_slice\_dict}(\mathcal{T}, r)$ \Comment{{\scriptsize For distributed TN}}\label{alg:line:distributed}
    \State $S \gets \texttt{create\_slice\_dicts}(B, P_r)$ List of slices of the \label{alg:line:create_slice_dicts}
    TN to be computed. Note: $|S| = |B|\cdot 4^{|P_r|}$.
    \State Distribute $S$ over MPI ranks
    \For{$s$ in a local subset of $S$}
        \State $\mathcal{T}' \gets \texttt{slice\_tn}(\mathcal{T}, s)$\label{alg:line:slice_tn}
        \State $D \gets \texttt{contract\_tn}(\mathcal{T'}, F)$ \Comment{{\scriptsize Calculate marginals}}\label{alg:line:calculate_marginals}
        \State $B \gets \texttt{extend\_bitstrings}(B, D)$. Use largest $K$ values, or other heuristics. \label{alg:line:extend_bts}
    \EndFor
    \State $F \gets \texttt{get\_next\_free}(F, M)$. Note: size of $F$ is a multiple of $M$.
\EndWhile
\Return $B$
\end{algorithmic}
\end{algorithm}

In \cref{alg:largest_sampling}, we present a procedure for approximately recovering Pauli terms with large magnitude coefficients from our backpropagated observable. Our algorithm works by calculating marginal distributions
over some subsets (batches) of qubits. We then select a
number of partial bitstrings (prefixes) from that marginal distribution with large coefficients magnitudes and use them as our
sample candidates. The algorithm proceeds by calculating a marginal for another
batch of qubits, conditioned on each previously sampled candidate. The marginal is used to create several longer candidate solutions by extending the previous candidate solution (Line~\ref{alg:line:extend_bts}).

We modify the sampling algorithm to run in a distributed setting, as shown in
Algorithm~\ref{alg:largest_sampling}. 
The algorithm consists of three key parts: creating a list of slice dictionaries (Lines~\ref{alg:line:distributed}-\ref{alg:line:create_slice_dicts}), in parallel over the slice dictionaries contracting the sliced tensor networks (Lines~\ref{alg:line:slice_tn}-\ref{alg:line:calculate_marginals}) and extending the set of bitstrings based on the marginals (Line~\ref{alg:line:extend_bts}).
Note that the same algorithm can be used for distributed random sampling. 
The difference in our case is Line~\ref{alg:line:extend_bts}, where we use
largest $K$ elements, instead of randomly sampling. Note that in random sampling, the strategy for number of samples is also different. This approach parallelizes over both TN slices (Line~\ref{alg:line:distributed}) \cite{lykov_step_dependent_2022, Simple_heuristic_Schutski_2020, classical_simulation_quantum_chen_2018} and over the candidate set of partial bitstrings which have been sampled (Line~\ref{alg:line:create_slice_dicts}).

\paragraph{Algorithm parameters}

There are two notable parameters in the algorithm: the batch size $M$, which
determines the size of the marginal distribution tensor, and the number of
samples to select at each round $K$. A higher $M$ results in a more expensive TN
contraction, while a smaller $M$ increases the number of sampling rounds $\lceil N/M\rceil$ and increases the variance of the procedure. We choose $M=9$ for our calculations. 

Adjusting the number of samples $K$ impacts the cost of simulation linearly,
as one needs to contract a separate TN for each bitstring candidate in $B$
(Line~\ref{alg:line:bitstrings}), and $|B| = K^{N/M}$ in the worst case.
However, the value $K$ need not be the same for each round, or for each call to Line~\ref{alg:line:extend_bts}. Thus, we can use value-based truncation based on the normalized marginal Pauli string coefficients and their cumulative weight. 
Specifically, we choose an error parameter $\alpha$ and select $K$ for each individual 
marginal distribution such that the highest $K$ marginals add up to $1-\alpha$
fraction of total probability weight.

\paragraph{Numerical results}

\begin{figure}
    \centering
    \includegraphics[width=1\linewidth]{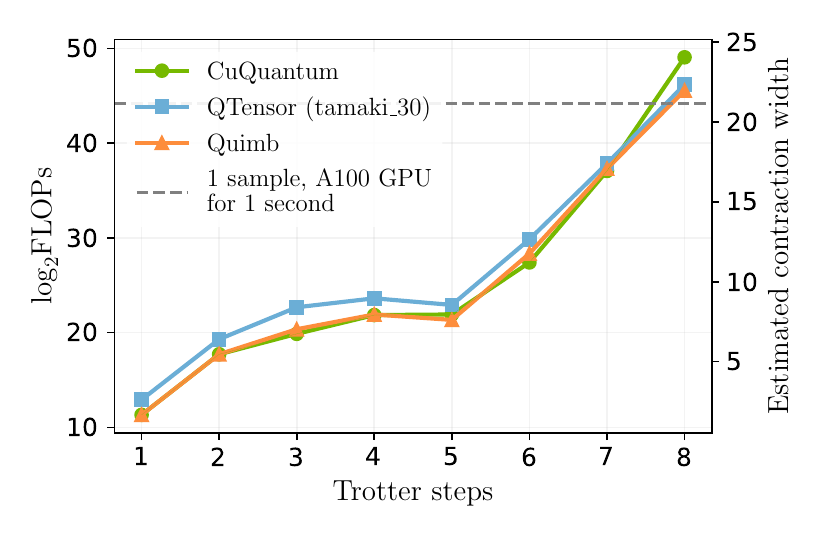}
    \caption{Estimated cost of evolving $Z_{62}$ using different
    tensor network contraction packages. The value of width is approximate, as
    different tradeoffs between
    width and FLOPs may be used by different packages.
    Note that contraction optimization time for different packages was
    different.
    The dotted line is based on perfect performance. Slicing is not performed for Quimb and QTensor.
    }
    \label{fig:tn_width}
\end{figure}

The approaches described above were used to simulate the Heisenberg-picture evolution
under the Hamiltonian in \cref{eq:Trotterized_evolution}.
In order to simulate the full polarization $M=\frac{1}{n}\sum_i Z_i$, we need to evaluate
each $Z_i$ individually.
In these numerics, we simulated a single $Z$ observable on qubit 62. We choose this qubit as 
it is close to the center of the connectivity map of the 127-qubit Eagle device.
This location results in a larger growth of the lightcone compared to qubits that are closer 
to edges. Thus, we can use the results for $Z_{62}$ as an upper bound on the simulation cost.

The limits of the TN-based simulation come from two factors: TN complexity and the number of
Pauli strings $|B|$. 
To contract any TN, one has to find a ``contraction tree''---an intermediate 
structure that determines the order in which tensors are contracted.
TN contraction complexity comes from the size of intermediate tensors, determined
by number of tensor indices. This number is often called ``contraction width'' due to the 
connection of finding the contraction tree to the tree decomposition problem.
Finding a good contraction tree is a challenging combinatorial optimization task~\cite{Markov_2008}.
If a TN contraction tree results in a high contraction width, one has to 
split the TN into ``slices", by finding some set of indices and fixing their values 
depending on slice index. This process may result in an exponential number TN slices, which
increases the time to solution.

To determine the feasible number of Trotter steps for simulation, we 
have to find the contraction trees prior to simulation. In practice,
since the first round of sampling has the smallest number of sliced indices,
it has the highest contraction width. We evaluate the performance of several state-of-the-art
TN contraction algorithms, as shown on~\cref{fig:tn_width}. 
Since the TN is set to produce a tensor with $M$ indices, the contraction width
is always larger than $M$. The final tensor is not considered an intermediary in~\cref{fig:tn_width}, but it affects the contraction path, 
which results in a constant scaling of with up to some Trotter step.

The contraction optimization was evaluated at default parameters for all packages, 
and for QTensor the Tamaki optimization algorithm was used with 30 seconds time
budget. Contraction optimization time varies significantly over packages.
CuQuantum is the fastest with sub-second times,
QTensor was time-limited to 30 seconds, and Quimb required in the order of 5 minutes.

Note that the number of elements in the largest intermediate scales with the width $w$ as $~4^w$.
In addition to the largest intermediate, other tensors require memory.
In practice, at width 14, it is feasible to run on one A100 GPU with 80 GB of memory.
Thus, for 7 Trotter steps, slicing is required and on average produces 16 slices per TN.

\begin{figure}
    \centering
    \includegraphics[width=1\linewidth]{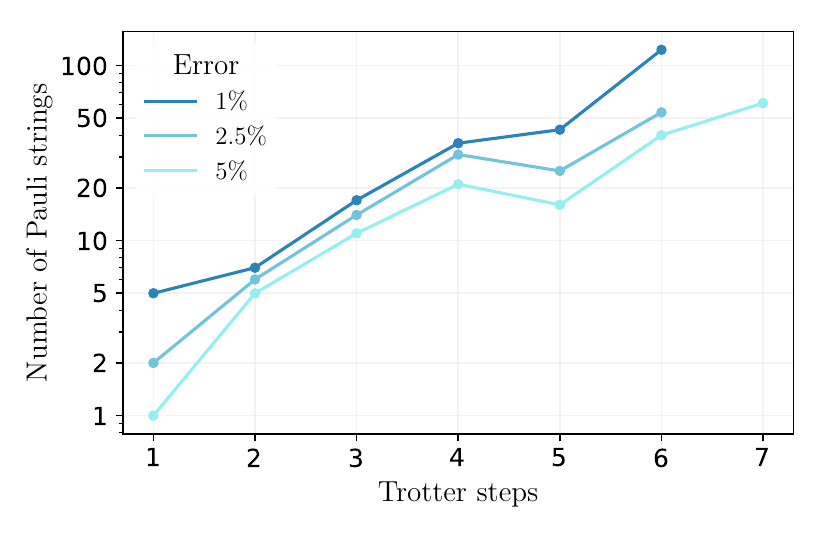}
    \caption{Number of Pauli strings in a backpropagated observable $Z_{62}$ to reach a
    fixed $L_2$ error threshold. We simulated 100 strings for 7 Trotter steps, 
    which was not enough to reach the 2.5\% threshold.
    }
    \label{fig:tn_num_paulis}
\end{figure}

The impact of the second factor, number of Pauli strings, is demonstrated 
in~\cref{fig:tn_num_paulis}. In our simulations, we use varying values of 
$\alpha$ to adjust the number of Pauli strings $K$ to balance
between the simulation time and the truncation error.

\begin{figure}
    \centering
    \includegraphics[width=1\linewidth]{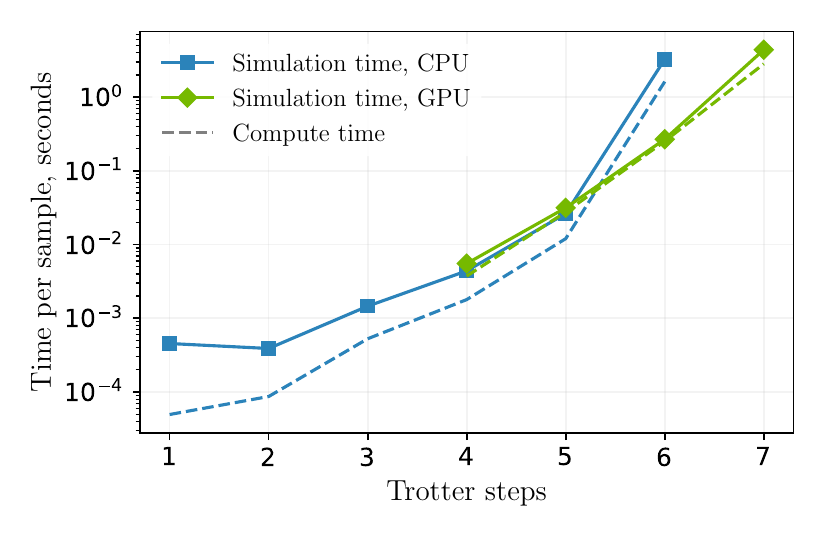}
    \caption{Contraction time per single Pauli string using the QTensor quantum
    circuit simulator.
    Dashed lines correspond to TN contraction time, while solid lines 
    correspond to total simulation time, including communication and 
    TN preparation, but excluding TN contraction tree optimization.
    }
    \label{fig:tn_time}
\end{figure}

The smaller simulations were run on a consumer-grade laptop CPU AMD Ryzen 9 5980HS.
For larger simulations that require more memory and compute, 
such as for 7 Trotter steps, we used up to 8 nodes of Argonne's Polaris supercomputer.
Each node of Polaris contains 4 NVIDIA A100 GPUs.
In both cases, we used the QTensor quantum circuit simulation package.
Simulation time is linearly proportional to the number of Pauli strings,
and one can adjust it based on desired quality.
\Cref{fig:tn_time} shows simulation time per single sample.
All experiments used the batch size $M=9$. Note that using GPU only benefits 
after the TN contraction involves large tensors, at 
TN width at least $ 12$. A similar behavior was observed for quantum circuit simulation~\cite{Lykov_GPU_QTensor}. Further improvements
to contraction performance are possible using the cuQuantum package~\cite{cuquantum_sdk}.

\section{Localization of the Trotter Error in the Polarization}\label{sec:localization}

In the main text, we considered a Trotterization of the XY model where the ordering of the terms preserves the polarization $M = \frac{1}{n}\sum_i Z_i$.
Instead, one may choose an ordering that breaks the symmetry, e.g. 
\begin{align}
U(\tau) = 
    \prod_{i,j\in \mathcal E(\Lambda)}
    e^{-iJ\tau Y_i Y_j}
    \prod_{i,j\in \mathcal E(\Lambda)}
    e^{-iJ\tau X_i X_j}\, .\label{eq:Trotterized_evolution_2}
\end{align}
In this case, the polarization $M$ may deviate from the initial value under the Trotterized dynamics.
Studying this deviation due to Trotter error may offer rich insights into the XY model, such as the existence of a phase transition between localization and quantum chaos~\cite{heyl_quantum_2019, sieberer_digital_2019}. 
In this section, we numerically and analytically explore the localization of the error in the Polarization $M$ in the Trotterized dynamics of the XY model.

\begin{figure}
    \centering
    \includegraphics[width=0.45\textwidth]{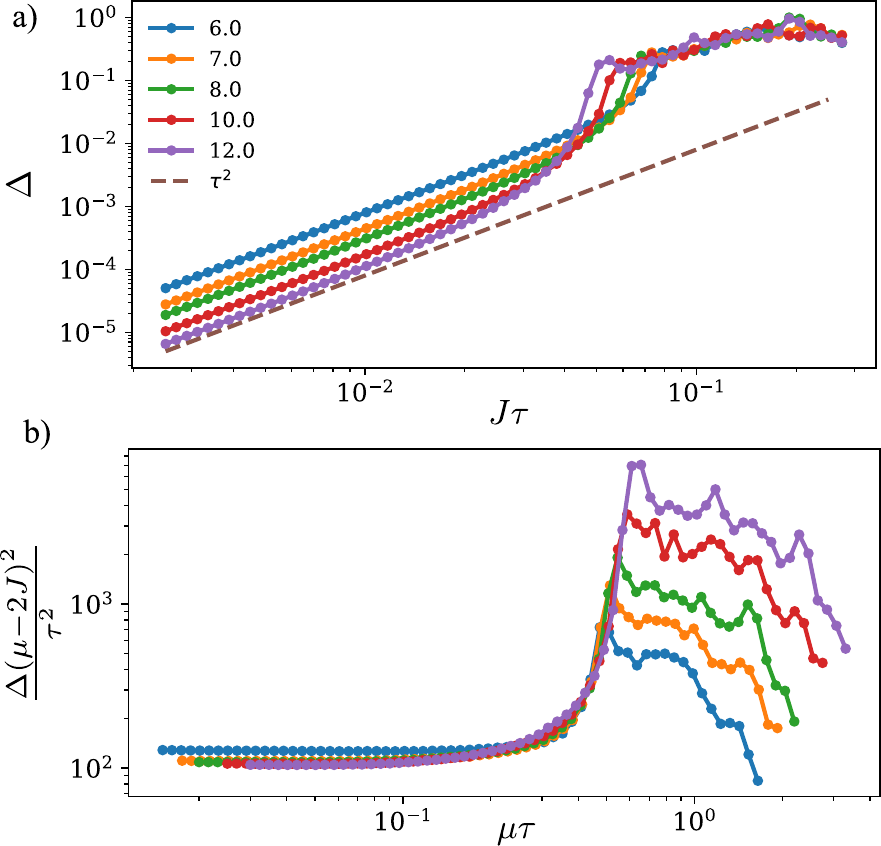}
    \caption{(a) Error of polarization at long time $\Delta$~[\cref{eq:Delta-def}] versus Trotter step size. Different lines correspond to different values of the local field strength $\mu$. The dashed line is a function scaling as $\tau^2$ for reference.
    (b) Data collapse highlights the stable regime where $\mu\tau < 1$. In this regime, we find that $\Delta\propto \tau^2/(\mu-J)^2$.}
    \label{fig:localization-numerics}
\end{figure}

Numerically, we consider the XY model in \cref{eq:XY-model} on a one-dimensional lattice of $n=12$ qubits with closed boundary conditions. 
We compute the long-time deviation 
\begin{align} 
	\Delta \equiv \lim_{t\rightarrow \infty} \abs{\avg{M}_t - \avg{M}_0}, \label{eq:Delta-def}
\end{align}
where $\avg{M}_0 = 1$ is the initial polarization given an initial state $\ket{\psi}$ is an eigenstate of $M$ and $\avg{M}_t \equiv \bra{\psi} [U(\tau)^\dag]^{t/\tau} M U(\tau)^{t/\tau}\ket{\psi}$ is the expectation value of the polarization in the state $\ket{\psi}$ evolved under the Trotterized dynamics in \cref{eq:Trotterized_evolution_2}. 
As shown in \cref{fig:localization-numerics}, for sufficiently small $\tau$, the polarization deviates from the conserved value by only a small error even in the long-time limit. 
The error can be made arbitrarily small, by either reducing the Trotter step $\tau$ or increasing the gap between different symmetry sectors by increasing $\mu$.
\Cref{fig:localization-numerics} suggests the existence of a quasi-conserved quantity $\tilde M$ whose overlap with $M$ is controlled by $\tau$ and $\mu$. \\

\textbf{Quasi-conserved quantity}---To argue for the existence of a quasi-conserved quantity, we first consider the Hamiltonian $H_f$ that generates the evolution over one Trotter step, i.e. 
\begin{align} 
	U(\tau)=e^{-iH_{f}\tau}.	 
\end{align}
To avoid crossing the branch cut in taking the logarithmic function of $U(\tau)$ (the $\pi$ quasi-energy), we additionally require $\tau\norm{H_f}\ll 1$. 
In the limit $\tau\to 0$, $\lim_{\tau\to0} U(\tau) = e^{-iH\tau}$ where $H$ is the exact Hamiltonian of the XY model in \cref{eq:XY-model}.
We define a differentiable path $H(s)$ with $H(0) = H$ and $H(\tau)=H_f$.
Using this differentiable path, we now extend the symmetry operator $M$ of $H(0)$ to the entire path $H(s)$ and subsequently the effective Hamiltonian $H_f = H(\tau)$. 
The conditions of the path $H(s)$ will be clear later.

\begin{figure}[b]
    \centering
    \includegraphics[width=0.25\textwidth]{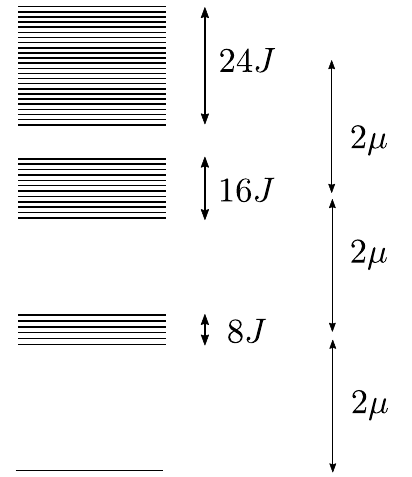}
    \caption{The energy spectrum of the XY model can be separated into different sectors corresponding to different values of the polarization. The width of each sector depends on the polarization and ranges from 0 for the $\ket{\uparrow}^N$ to $4n J$. These sectors are gapped from each other if $\mu$ is large enough.}
    \label{fig:energy-spectrum}
\end{figure}

Assume at some $s$, $[M(s),H(s)]=0$, we will find the derivative of $M(s)$ such that 
\begin{align} 
	 [M'(s),H(s)]+[M(s),H'(s)]=0.\label{eq:cond-for-M'}
\end{align}
This condition implies $[M(s+ds), H(s+ds)]=0$.
Since $M(s)$ and $H(s)$ commute, there is a basis $\{\ket{j(s)}\}$ such that $H(s)\ket{j(s)}=\epsilon_j(s)\ket{j(s)}$ and $M(s)\ket{j(s)}=m_j(s)\ket{j(s)}$, where $\epsilon_j(s)$ and $m_j(s)$ are the corresponding eigenvalues.
In this basis, \cref{eq:cond-for-M'} is equivalent to
\begin{equation}
    M'(s)_{jk}[\epsilon_k(s)-\epsilon_j(s)] = H'(s)_{jk}[m_k(s)-m_j(s)], 
\end{equation}
which implies
\begin{align} 
	M'(s)_{jk} = \frac{m_j(s)-m_k(s)}{\epsilon_j(s)-\epsilon_k(s)} H'(s)_{jk}. \label{eq:M'-def}
\end{align}
The construction of $M'(s)$ is well-defined if  $m_j(s) = m_k(s) $ for any $\epsilon_j(s) = \epsilon_k(s)$, in which case $M'(s)_{jk}=0$.
To make sure this condition holds at $s = 0$ for the XY model, i.e. $\epsilon_j(0)=\epsilon_k(0)$ implies $m_j(0) = m_k(0)$, we assume that $\mu \geq (4 n + 2) J$~(\cref{fig:energy-spectrum}).
Note that this condition can be relaxed to $\mu \gtrsim 2J$ if the initial state $\ket{\psi(0)}$ is in the low-energy subspace.
Additionally, the condition $\mu \geq (4 n + 2)J$ ensures that the energy of different symmetry sectors are also gapped from each other, i.e. $\abs{\epsilon_j(s)-\epsilon_k(s)}\geq 2\mu-(8n+4)J$ if $m_j(s)\neq m_k(s)$.

To see that \cref{eq:M'-def} is well defined at all $s\leq \tau$, we first notice that if $\tau$ is sufficiently small, the gap between the energy of different symmetry sectors remains.
Therefore, $\epsilon_j(s) = \epsilon_k(s)$ only if $\ket{j(s)}$ and $\ket{k(s)}$ belong to the same symmetry sector.
We now prove that if $m_j(s) = m_k(s)$ at a some $s$, then at $s+ds$, $m_j(s+ds) = m_k(s+ds)$. 
By induction, it implies that $m_j^{s}$ remains the same for all eigenstates within the same symmetry sector at all $s$.
This guarantees that \cref{eq:M'-def} is well defined for small enough $\tau$ such that the energy gap between the symmetry sectors remains open.

Given that the eigenstate $\ket{j(s)}$ may be degenerate, under the perturbation theory, we can write the eigenstate at $s+ds$ as
\begin{align}
    &\ket{j(s+ds)} = \sum_{\epsilon_k(s) = \epsilon_j(s)} c_k \ket{k(s)} \nonumber\\
    &\quad+ ds \sum_{\epsilon_q(s) \ne \epsilon_j(s)} \frac{H'(s)_{qj}}{\epsilon_q(s) - \epsilon_j(s)} \ket{q(s)},
\end{align}
where $c_k$ are some coefficients.
The normalization of $\ket{j(s+ds)}$ implies that $\sum_{k} \abs{c_k}^2 = 1 - \O{ds^2}$.
We have
\begin{widetext}
\begin{align} 
	 m(s+ds)_j
	 &=\bra{j(s+ds)}M(s+ds)\ket{j(s+ds)} \\
	 &= m_j(s) \sum_{\epsilon_k(s) = \epsilon_j(s)} |c_k|^2 
	 + ds \sum_{\epsilon_k(s)= \epsilon_{k'}(s) = \epsilon_j(s)}\bra{k'(s)} M'(s) \ket{k(s)}\nonumber\\
	&\quad\quad\quad\quad\quad +  ds \sum_{\epsilon_k(s) = \epsilon_j(s)}\sum_{\epsilon_q(s) \ne \epsilon_j(s)}
	 \frac{H'(s)_{qj}}{\epsilon_q(s) - \epsilon_j(s)} \bra{k(s)} M(s)\ket{q(s)} + \text{h.c.} + \O{ds^2}\label{eq:derivative-of-m2}\\
	 & = m_j(s) + \O{ds^2}.\label{eq:derivative-of-m}
\end{align}
\end{widetext}
Here, the second term in \cref{eq:derivative-of-m2} vanishes because $m_k(s)=m_{k'}(s)$ and, per \cref{eq:M'-def}, $M'(s)_{kk'} = 0$.
The third term and its Hermitian conjugate in \cref{eq:derivative-of-m2} both vanish because $\ket{k(s)}$ and $\ket{q(s)}$ belong to different symmetry sectors and are orthogonal to each other. 
Since there is no $\mathcal{O}(ds)$ in \cref{eq:derivative-of-m}, it implies that $\frac{d}{ds}m_j(s) = 0$.
This condition describes a path where symmetry sectors (defined by eigenstates sharing the same $m$) do not cross each other and eigenvalues crossing is allowed within each sector. \\

\textbf{Trotter error at long-time}---We have shown that for sufficiently small $\tau$, there exist an effective conserved quantity whose expectation value is constant in time. In practice, we do not know the exact form of this conserved quantity and only measure the original operator. This results in a Trotter error that is constant in  the total simulation time $t$ but scales with $\tau$.

We limit the error analysis to the first order expansion of $\tau$ in $H_f$, i.e. $H(s) = H + sV$ where $V$ contains commutators between the terms of the Hamiltonian $H$~\cite{childs_theory_2021}. 
Then
\begin{equation}
    M'_{jk} = \frac{(M(\tau)-M(0))_{jk}}{\tau} = \frac{m_j(\tau) - m_k(\tau)}{\epsilon_j(\tau)-\epsilon_k(\tau)} V_{jk}. \label{eq:B10}
\end{equation}
At $t=0$, the initial state $\ket{\psi}$ is an eigenstate of $H(0)$ and $M(0)$ satisfying $H(0)\ket{\psi}=\epsilon_j(0) \ket{\psi}$ and $M(0)\ket{\psi}=m_j(0)\ket {\psi}$, where $j$ is the index of the corresponding eigenvalue. 
Expanding $\ket{\psi}$ in the eigenstates of $H(\tau)=H_f$, we have
    \begin{equation}
        \ket{\psi}  = \sum_{m_k = m_j} c_{k}\ket{k(s)} 
        + \sum_{m_q\neq m_j}c_q\ket{q(s)},
    \end{equation}

where $c_{k}\sim\mathcal{O}(1)$ runs over eigentates in the same sector as $\ket{\psi}$ and $c_{q}$ runs over the other eigenstates.
The perturbation theory implies $c_q\sim\mathcal{O}(\tau/\lambda)$ with $\lambda \sim \O{\mu}$ being the gap between symmetry sectors of $H(s)$. For normalization, $\sum_k|c_k|^2 = 1 - \mathcal{O}(\tau^2/\lambda^2)$. Since $M(\tau)$ is conserved, its expectation value at time $t$ that is an integer multiple of $\tau$ is
\begin{align}
    \bra{\psi(t)}M(\tau)\ket{\psi(t)} 
    &= m_j(\tau)   \sum_{\epsilon_k = \epsilon_j} |c_{k}|^2 +\sum_{\epsilon_q\neq \epsilon_j} |c_{q}|^2 m_q(\tau)\nonumber\\
    & = m_j + \O{\tau^2/\lambda^2},
\end{align}
which is independent of $t$.
To show that $\bra{\psi(t)}M(0)\ket{\psi(t)}$ is also independent of the total time $t$, we use \cref{eq:B10}
and
\begin{align}
    \bra{\psi(t)}\tau M'(\tau)\ket{\psi(t)} 
    = &\tau \sum_{\substack{\epsilon_k = \epsilon_j\\
    \epsilon_q\neq \epsilon_j}} c_{k}^*c_{q} e^{it(\epsilon_k(\tau) - \epsilon_q(\tau))} M'_{k,q} \nonumber\\
    &+ \text{h.c.} +\mathcal{O}(\tau^3).
\end{align}
The above quantity does not have $\mathcal{O}(\tau)$ because $M'$ does not couple states within the same sector by construction. The first term on the right-hand side scales as $\sim \mathcal{O}(\tau^2/\lambda)$ and has a fast fluctuation due to the gap $\abs{\epsilon_k(\tau)- \epsilon_q(\tau)}\sim \lambda$.
In particular, the time average of this oscillation vanishes in the long-time limit. Therefore, at a long time $t$, \cref{eq:B10} implies
\begin{align}
&\bra{\psi(t)}M(0)\ket{\psi(t)}\nonumber\\
&= \bra{\psi(t)}M(\tau)\ket{\psi(t)}
- \tau \bra{\psi(t)}M'(\tau)\ket{\psi(t)} \nonumber\\
&= m_j + \mathcal{O}(\tau^2/\lambda^2).  
\end{align}
In other words, the initial polarization $M$ is also approximately conserved up to a correction $\mathcal{O}(\tau^2/\lambda^2)$ that is independent of $t$.
Comparing with our numerical results shown in \Cref{fig:localization-numerics}(b), the error agrees with the scaling $\tau^2/(\mu-2J)^2$ at small $\tau$ (note that $\lambda = \mu - 2J$ in this numerics). \Cref{fig:localization-numerics}(b) also shows the breakdown of the Trotter approximation when $\mu\tau \sim \mathcal{O}(1)$.

\section{OBP Calculation details }\label{sec:numerics_details}

\begin{table*}[tph]
\caption{75-qubit Operator Backpropagation Numerics}
\centering
\begin{tabular}{ |P{1.1cm}||P{1cm}|P{1cm}|P{1cm}||P{1.2cm}|P{1.2cm}||P{1.2cm}|P{1.2cm}||P{1.4cm}|P{1.2cm}||P{1.2cm}|P{1.2cm}||P{1cm}|  }
 \hline
 OBP Trotter Steps & Initial $L_2$ Budget & Final $L_2$ Budget & Total $L_2$ Budget & Initial OBP runtime & Final OBP runtime & Unique Paulis (initial)& Unique Paulis (final) & Mean Paulis per $Z_i$ (initial)& Mean Paulis per $Z_i$ (final)& Median Paulis per $Z_i$ (initial)& Median Paulis per $Z_i$ (final) & QWC Pauli Groups \\
 \hline

1 & 0.0 & 0.01 & 0.01 & 0.29 & 0.095 & 1888 & 222 & 54.24 & 3.9067 & 56 & 4 & 3\\ 
2 & 0.0 & 0.01 & 0.01 & 0.598 & 0.61 & 14043 & 223 & 428.4533 & 4.9467 & 452 & 5 & 3\\ 
3 & 0.001 & 0.009 & 0.01 & 1.059 & 0.089 & 795 & 297 & 49.0267 & 7.84 & 51 & 8 & 6\\ 
4 & 0.001 & 0.009 & 0.01 & 1.885 & 0.125 & 933 & 369 & 71.36 & 9.8133 & 75 & 10 & 6\\ 
5 & 0.001 & 0.009 & 0.01 & 3.344 & 0.152 & 1003 & 370 & 87.9067 & 14.5867 & 93 & 15 & 8\\

\hline
\end{tabular}
\label{Tab:obp_numerics_75}
\end{table*}

\begin{table*}[tph]
\caption{127-qubit Operator Backpropagation Numerics}
\centering
\begin{tabular}{ |P{1.1cm}||P{1cm}|P{1cm}|P{1cm}||P{1.2cm}|P{1.2cm}||P{1.2cm}|P{1.2cm}||P{1.4cm}|P{1.2cm}||P{1.2cm}|P{1.2cm}||P{1cm}|  }
 \hline
 OBP Trotter Steps & Initial $L_2$ Budget & Final $L_2$ Budget & Total $L_2$ Budget & Initial OBP runtime & Final OBP runtime & Unique Paulis (initial)& Unique Paulis (final) & Mean Paulis per $Z_i$ (initial)& Mean Paulis per $Z_i$ (final)& Median Paulis per $Z_i$ (initial)& Median Paulis per $Z_i$ (final) & QWC Pauli Groups \\
 \hline
1 & 0.0 & 0.01 & 0.01 & 0.964 & 1.228 & 18210 & 410 & 301.244 & 4.448 & 159 & 4 & 5\\ 
2 & 0.0 & 0.01 & 0.01 & 20.364 & 92.994 & 1525084 & 415 & 23156.417 & 5.535 & 9642 & 5 & 5\\ 
3 & 0.0025 & 0.0075 & 0.01 & 24.105 & 0.835 & 7985 & 809 & 221.968 & 10.937 & 271 & 9 & 8\\ 
4 & 0.005 & 0.02 & 0.025 & 174.489 & 1.554 & 15826 & 809 & 425.952 & 11.85 & 506 & 11 & 8\\ 
5 & 0.005 & 0.02 & 0.025 & 1770.817 & 5.293 & 58066 & 881 & 1435.204 & 19.44 & 1707 & 20 & 8\\
 \hline
\end{tabular}
\label{Tab:obp_numerics_127}
\end{table*}

In this section, we discuss the details and overhead of the operator backpropagation calculations which were performed using the OBP Qiskit Addon~\cite{qiskit-addon-obp} in order to generate \cref{fig:hw_OBP_vs_no_OBP}. All of the details discussed here are presented in \cref{Tab:obp_numerics_75} and \cref{Tab:obp_numerics_127}.  For both the 75 and 127 qubit experiments, the operator $M = \frac{1}{n}\sum_i Z_i$ was estimated by backpropagating each $Z_i$ in two phases which we will refer to as the initial OBP step and final truncation step. The initial OBP step is performed by partitioning the target circuit into slices which contain a single commuting layer of generalized two qubit gates and then backpropagating each $Z_i$ with an $L_2$ error budget that is evenly divided across all slices. The final truncation step then takes this backpropagated observable and performs another round of truncation using a separate $L_2$ error budget. The initial, final, and total $L_2$ error budgets for each computation are reported in Tables \ref{Tab:obp_numerics_75} and \ref{Tab:obp_numerics_127}. 

If one had unbounded classical resources and time, one would obtain the smallest backpropagated operator size by performing the initial OBP step with no error budget and instead allocating one's entire $L_2$ error budget to the final truncation step. This is a consequence of observing that unitary transformations preserve the $L_2$ norm, truncated observable terms after different slices during OBP need not be orthogonal, and thus that errors incurred from truncation after different slices must be summed using the triangle inequality rather than the Euclidean norm. Thus the $L_2$ error budget we choose to allocate to the initial OBP step serves to reduce the operator size during backpropagation, limiting classical overhead, while at the same time resulting in an increase in the final operator size. This effect can be seen in Tables \ref{Tab:obp_numerics_75} and \ref{Tab:obp_numerics_127} by inspecting the rows corresponding to 2 backpropagated Trotter steps. Here we observe that after the initial OBP step with 0 $L_2$ error budget, there are 1,525,084 unique Pauli operators across all backpropagated observables, which is two orders of magnitude larger than we see after backpropagating through 5 Trotter steps with an initial $L_2$ budget of 0.001. However, after the final $L_2$ truncation, we see the number of unique Paulis drop dramatically, as we expect. For all OBP calculations, the initial and final error budgets were chosen to minimize the total number of qubit-wise commuting groups which contain all unique Pauli operators, while also limiting the classical overhead of the OBP procedure.

\section{MPS Calculations}\label{sec:mps_simulations}

In \cref{fig:hw_OBP_vs_no_OBP}b and \cref{fig:hw_OBP_vs_no_OBP}d we compare expectation values computed on a quantum processor using OBP with those obtained via a matrix product state (MPS) tensor network. All MPS calculations were performed by converting qiskit QuantumCircuit objects to quimb MPS objects using the qiskit-quimb and qiskit-addon-AQC-Tensor packages \cite{qiskit-quimb, qiskit-addon-aqc-tensor, Gray2018}. This conversion truncates all eigenvalues of the MPS below a threshold of $|\lambda| \le 10^{-10}$.

\section{Error Mitigation}\label{sec:error_mitigation}

All experiments were error-mitigated using zero-noise extrapolation (ZNE)~\cite{temme_error_2017,li_efficient_2017,kim_scalable_2023,kandala_error_2019} via probabilistic error amplification (PEA)~\cite{li_efficient_2017,mcdonough_automated_2022,mari_extending_2021,ferracin_efficiently_2024, Endo_2018} as well as twirled readout error extinction (TREX)~\cite{van_den_berg_model-free_2022}. For all experiments TREX was calibrated using a total of $32768$ twirling samples with $32768$ shots per sample. 
The noise learning and noise amplification procedures used to apply PEA were performed as in ~\cite{kim2023evidence}.
All twirling configurations sampled during PEA were executed with $64$ shots, noise amplification factors of $[1, 1.5, 2.25,3]$ were used for PEA, and these noisy expectation values were extrapolated using a hybrid of exponential and linear fits. This hybrid fitting consisted of first fitting results using an exponential model, and then rejecting outcomes which were highly non-physical or had severe fitting error in favor of a linear fit. This hybrid scheme is motivated by the fact that when extrapolating expectation values near zero, the statistical error of each estimate is comparable to the estimate's magnitude and exponential fitting can produce estimates which are orders of magnitude beyond physical values. Thus, this hybrid extrapolation scheme is designed to prioritize exponential extrapolation while also filtering outlier data points.

\section{Device Specification}\label{sec:device_spec}

\begin{figure*}
    \centering
    \includegraphics[width=0.85\linewidth]{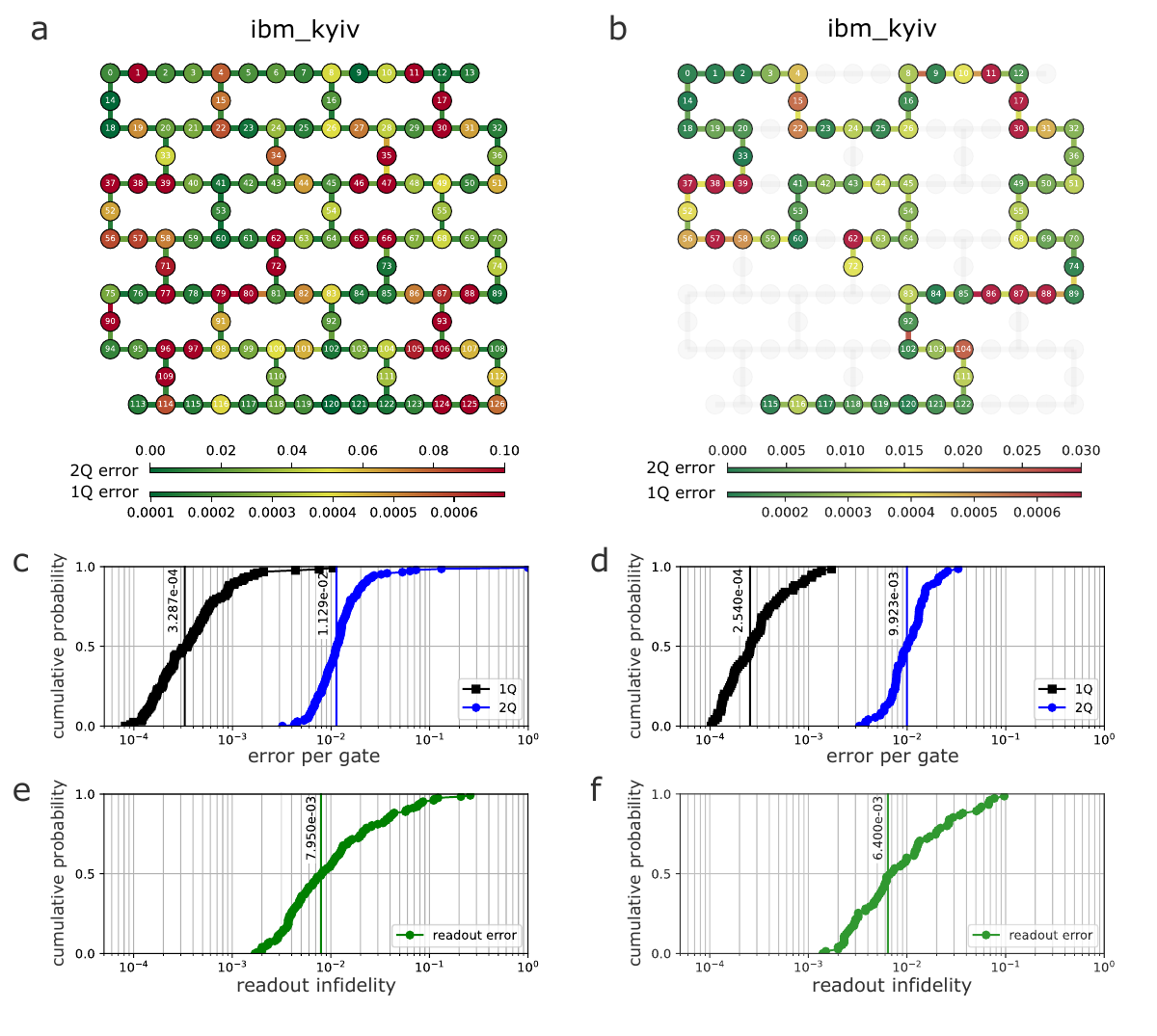}
    \caption{
        \textbf{Summary of single/two qubit gate errors and readout fidelity distribution on} \texttt{ibm\_kyiv}. a (b) Single/two qubit error per gate (EPG) values shown for the 127 (75) qubit experiments on \texttt{ibm\_kyiv}. c (d) Single/two qubit EPG distribution of device subsets for 127 (75) qubit experiments on \texttt{ibm-kyiv}. Median value indicated in plots with a vertical line. Mean value of single qubit EPG are $6.17 \times 10^{-4} \pm  1.20 \times 10^{-3} \ (3.64 \times 10^{-4} \pm  3.17 \times 10^{-4})$ and mean values of two qubit EPG are $2.09 \times 10^{-2} \pm  8.3 \times 10^{-2} \ (1.12 \times 10^{-2} \pm  5.47 \times 10^{-3})$. e (f) Readout fidelity distribution of device subsets for 127 (75) qubit experiments on \texttt{ibm\_kyiv}. Median values indicated in plots with vertical lines, mean values are $2.11 \times 10^{-2} \pm  3.62 \times 10^{-2} \ (1.57 \times 10^{-2} \pm  2.01 \times 10^{-2})$.}
    \label{fig:combined_device_plot}
\end{figure*}

The experimental results presented in \cref{sec:experiment} were obtained on IBM Quantum's {\texttt{ibm\_kyiv}} QPU, which is an Eagle-series quantum processor consisting of fixed-frequency transmon qubits with capacitive coupling between 127 neighboring qubits arranged in a heavy-hexagonal lattice. The 127 and 75 qubit experiments we present in \cref{sec:experiment} were executed on different days. In \cref{fig:combined_device_plot} we show the qubit layouts used for the 127 and 75 qubit experiments where the nodes and edges are colored to indicate the single and two qubit error per gate (EPG). The subplots below show the distribution of single and two qubit gate errors as well as the readout infidelity of the qubits used in each experiment. In table \cref{Tab:kyiv_qubit_properties} we report statistics of the $T_1$ and $T_2$ values for the subsets of {\texttt{ibm\_kyiv}} which were used for the 127 and 75 qubit experiments. 

\begin{table*}[tph]
\centering
\begin{tabular}{ ||P{1.5cm}|P{1.2cm}|P{2.5cm}|P{1.1cm}| P{1.1cm} ||P{1.2cm}|P{2.5cm}|P{1.1cm}| P{1.1cm} ||}
 \hline
 & \multicolumn{4}{|c||}{ibm\_kyiv (127Q)} & \multicolumn{4}{|c|}{ibm\_kyiv (75Q)} \\
 \hline\hline
  & median & mean & min & max & median & mean & min & max\\ 
 \hline
$T_1\ (\mu s)$ & 256.13	& 256.82 ± 99.61 &	15.51	&	494.88 & 265.56	& 271.85 ± 82.21 & 	123.16 &	 507.41 \\ 
$T_2\ (\mu s)$ & 112.17	& 142.96 ± 111.37 &	6.89	&	568.76 & 142.49	& 168.84 ± 120.87 & 	19.13 &	 530.91 \\ 
 \hline
\end{tabular}
\caption{\textbf{Summary of single qubit properties on \texttt{ibm\_kyiv}}. Reported $T_1$ and $T_2$ were obtained from daily calibration data, reported via the IBM Quantum Platform}
\label{Tab:kyiv_qubit_properties}
\end{table*}

\FloatBarrier

\bibliography{main}
\end{document}